%% file: TGIFA_arXiv.tex
\documentclass[12pt]{article}
\pdfoutput=1
\usepackage{authblk}

\input{preamble} 	
\begin{document}

\title{Missing data imputation using a \\truncated Gaussian infinite factor model \\with application to metabolomics data}
\author[1]{Kate Finucane}
\author[2,3]{Lorraine Brennan}
\author[4]{Roberta De Vito}
\author[5]{Massimiliano Russo}
\author[1]{Isobel Claire Gormley\thanks{claire.gormley@ucd.ie}}

\affil[1]{\footnotesize{School of Mathematics and Statistics,  University College Dublin, Ireland.}}
\affil[2]{\footnotesize{School of Agriculture and Food Science, University College Dublin, Ireland.}}
\affil[3]{\footnotesize{Conway Institute of Biomolecular and Biomedical Research, University College Dublin, Ireland.}}
\affil[4]{\footnotesize{Dipartimento di Scienze Statistiche, Sapienza University of Rome, Rome, Italy.}}
\affil[5]{\footnotesize{Department of Statistics, The Ohio State University, Columbus, OH, USA.}}

\date{}

\maketitle

\begin{abstract}

Metabolomics is the study of small molecules in biological samples. Metabolomics data are typically high-dimensional and contain highly correlated variables and frequent missing values. Both missing at random (MAR) data, due to acquisition or processing errors, and missing not at random (MNAR) data, caused by values falling below detection thresholds, are common. Thus, imputation is a critical component of downstream analysis. Existing imputation methods generally assume one type of data missingness mechanism, or impute values outside the data's physical constraints.

A novel truncated Gaussian infinite factor analysis (TGIFA) model is proposed to perform statistically principled and physically realistic imputation in metabolomics data. By incorporating truncated Gaussian assumptions, TGIFA respects the data's physical constraints, while leveraging an infinite latent factor framework to capture high-dimensional dependencies without pre-specifying the number of latent factors. Our Bayesian inference approach enables uncertainty quantification in both the values of the imputed data, and the missing data mechanism. A computationally efficient exchange algorithm enables scalable posterior inference via Markov Chain Monte Carlo. We validate TGIFA through a comprehensive simulation study and demonstrate its utility in a motivating urinary metabolomics dataset, where it yields useful imputations, with associated uncertainty quantification. Open-source R code, available at \href{https://github.com/kfinucane/TGIFA}{github.com/kfinucane/TGIFA}, accompanies TGIFA.

\end{abstract}

\maketitle

\section{Introduction}

Metabolomics is the study of small molecules in biological samples, referred to as metabolites. The metabolome is the complement of metabolites in a sample, which can reveal information about altered metabolic pathways when examined under different conditions \citep{Kosmides2013, Zhong2022}. Applications of metabolomics span from disease biomarker discovery \citep{Tounta2021} to food and nutrition research \citep{LeVatte2021}, with data primarily acquired through nuclear magnetic resonance spectroscopy and mass spectrometry \citep{Spicer2017}.

From a statistical perspective, metabolomics data pose several challenges. Metabolomics studies typically generate data in which the number of observations, $n$, is much smaller than the number of variables, $p$, which presents a challenge for many statistical modelling tools \citep{worley2013, Zhao2019}. Further, the variables in metabolomics data are often highly correlated with each other (multicollinearity), again posing challenges for many statistical models \citep{Blaise2016}. Many tools exist to aid in the processing and analysis of such data \citep{Wishart2022, Pang2024} and dimension reduction techniques are often utilised to address the issues of $n \ll p$ and multicollinearity \citep{Worheide2021}. Factor analysis is one such technique \citep{Liland2011, Meng2016, murphy2020} modelling the covariance structure of many observed variables via a smaller set of underlying latent factors. Although metabolomics data are comprehensive and high-dimensional, missing data are a prevalent feature \citep{Taylor2021, Sun2024}. Two types of missingness mechanism are typical in metabolomics data: missing at random (MAR) data, which arise when a metabolite is present but is undetected due to technical or processing errors, and missing not at random (MNAR) data, which arise when either a metabolite is not present or it is present but at a concentration below the limit of detection (LOD) \citep{Wilson2022}.

Many methods exist for missing data imputation in general, including fixed-value imputation methods (e.g., variable-specific mean, median, half-minimum, or minimum-value imputation) \citep{Gromski2014, Wei2018, Sun2024}, $k$-nearest-neighbours-truncation (KNN-TN) \citep{Shah2017}, imputation using singular value decomposition (SVD) \citep{TrevorHastie2021}, left-censored missing value imputation approaches, such as GSimp \citep{Wei2018a}, Bayesian principal component analysis (BPCA) \citep{Oba2003}, and random forest (RF) models, e.g., \cite{Stekhoven2012}. However, most existing methods fail to accommodate both the MAR and MNAR mechanisms simultaneously or impute values outside of the physical constraints of the data.

To ensure imputation of physically meaningful values and cognisance of different missing data types, while accounting for the multicollinearity and $n \ll p$ characteristics of metabolomics spectrometry data, we introduce a novel Truncated Gaussian Infinite Factor Analysis (TGIFA) model. Under TGIFA, missing data are imputed under the assumption of a truncated Gaussian factor analysis model, thereby respecting the physical constraints of the data and inducing parsimony given the $n\ll p$ setting. Additionally, latent shrinkage priors facilitate an infinite factor model \citep{Bhattacharya2011, Shah2019}
obviating the need to select the number of latent factors underlying the $p$ correlated variables. Importantly, given the presence of multiple types of missingness mechanism in metabolomics data, TGIFA allows for imputation of MAR and MNAR data appropriately and simultaneously. In contrast to imputation methods which lack a probabilistic basis, TGIFA's underpinning probabilistic model and inference in the Bayesian framework naturally allow uncertainty quantification of the imputed values, and their missingness type. Though the truncated Gaussian distribution is useful when modelling data that have a restricted domain, its use can bring computational difficulties, especially in high-dimensional settings due to the requirement to evaluate high-dimensional integrals. Inspired by its use in intractable likelihood settings (e.g., \cite{Piancastelli2024}), TGIFA employs an exchange algorithm \citep{murray2006} to eliminate the need to evaluate such integrals and facilitate computationally efficient inference when using a truncated multivariate Gaussian distribution. Thus, TGIFA allows for missing data to be imputed in a statistically and physically principled manner, with inherent uncertainties in the imputed values quantified.

In what follows, Section \ref{section:broccoli} introduces a urinary metabolomics dataset which motivates the TGIFA model. Interest lies in imputing the missing data in a manner which is cognisant of both MAR and MNAR missingness types, respects the data's multicollinearity and physically-enforced non-negative support, and quantifies the uncertainty associated with each imputed value. Section \ref{section:methods} outlines the TGIFA model, while Section \ref{section:inf} describes its inference using the exchange algorithm and provides details on imputation. Section \ref{section:simulation} details a simulation study performed to assess the performance of TGIFA and in Section \ref{section:real} TGIFA is used to impute missing values in the urinary metabolomics dataset. Section \ref{section:discussion} concludes with a discussion. 
To facilitate widespread use of TGIFA, associated R code is available at \href{https://github.com/kfinucane/TGIFA}{github.com/kfinucane/TGIFA}, with which all results herein were produced.

\section{A urinary metabolomics dataset}
\label{section:broccoli}

A typical urinary metabolomics dataset motivates the proposed TGIFA approach. The dataset is derived from a liquid chromatography mass spectrometry (LC-MS) study examining the postprandial response of the metabolome (the complement of metabolites in a sample) to broccoli consumption \citep{McNamara2023}. The study included $18$ participants, pre and post-consumption of cooked broccoli. The LC-MS data used in the current work are from the baseline pre-consumption samples and include $n = 18$ participants and $2032$ variables. Ethical approval was granted by the UCD Sciences Human Research Ethics Committee (LS-15-69-Brennan). All participants provided written informed consent.

\begin{figure}[htp]
\begin{subfigure}{\textwidth}
  \centering
  \includegraphics[width=\textwidth, keepaspectratio]{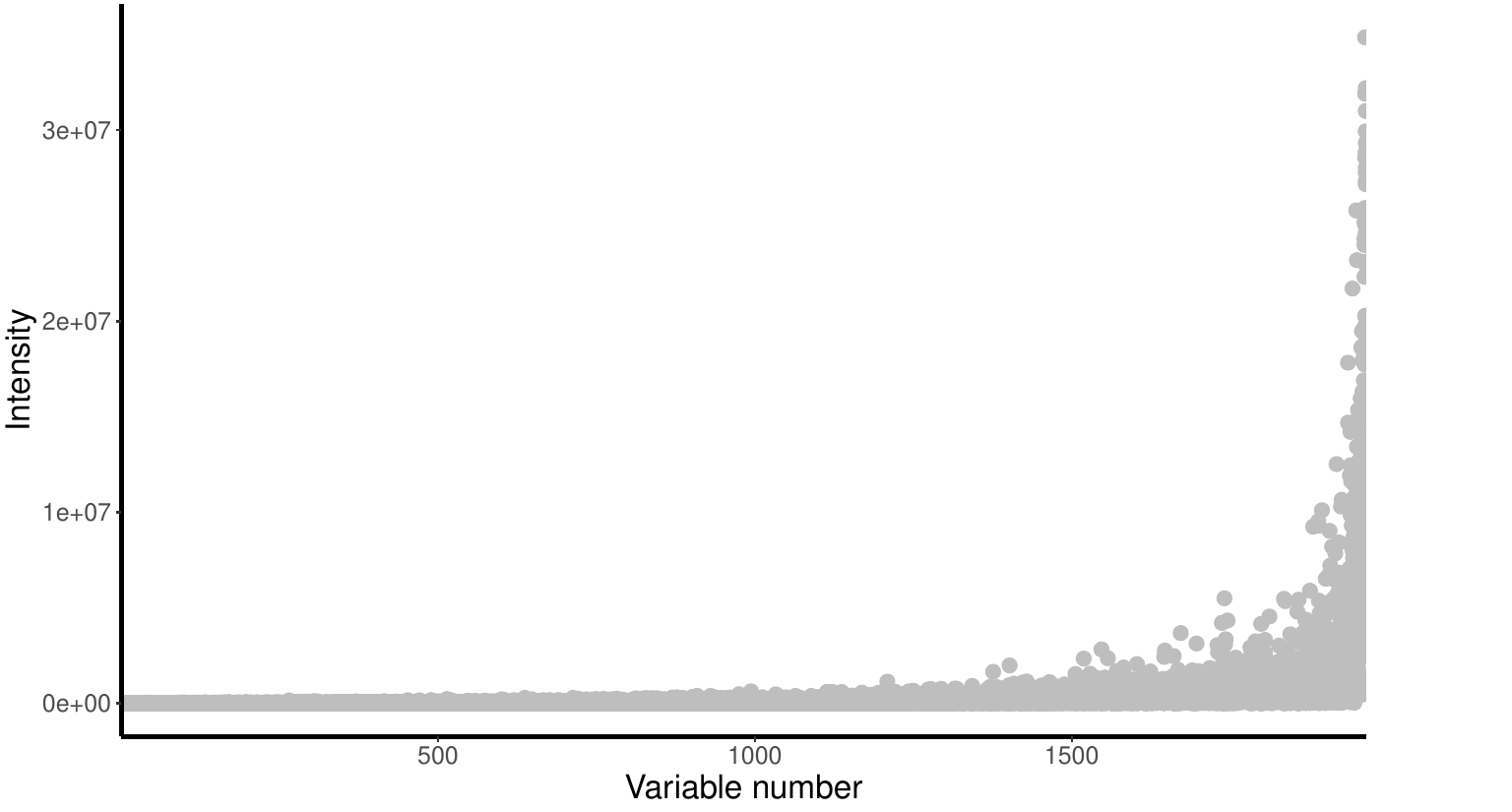}
  \caption{}
  \label{fig:broc_features_full_data_sec}
\end{subfigure}
\begin{subfigure}{\textwidth}
  \centering
  \includegraphics[width=\textwidth, keepaspectratio]{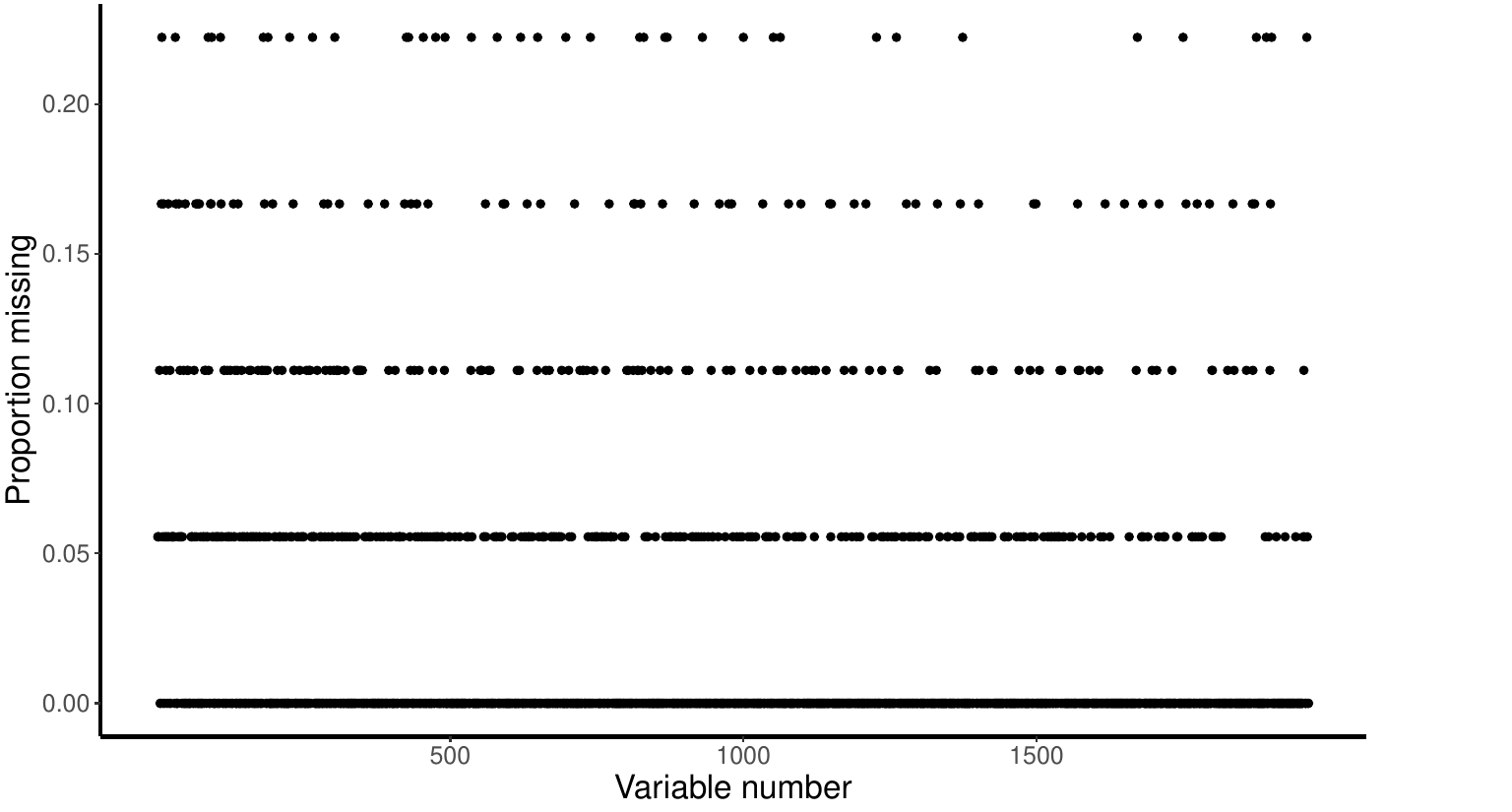}
  \caption{}
  \label{fig:broc_features_missprop_data_sec}
\end{subfigure}
\caption{(a) The untargeted urinary metabolomics data from baseline (pre-consumption) samples. (b) The proportion of missingness per variable after preprocessing.}
\label{fig:broc_missing_prop}
\end{figure}

As removing variables with $>20\%$ missingness is typical in metabolomics research \citep{Bijlsma2006}, here a total of 68 variables, that had a higher $>25\%$ of their entries missing, were removed prior to analysis. The resulting dataset consists of $p = 1964$ variables, with an overall missingness rate of 2.69\%. There are 1391 variables with no missing entries, and the mean missingness proportion across the 573 variables with missing entries is 9.22\%. Figure \ref{fig:broc_missing_prop} provides an overview of the final dataset.

\section{Bayesian truncated Gaussian infinite factor analysis}
\label{section:methods}

Due to their parsimony, factor analysis models are widely used approaches for modelling high-dimensional, multicollinear data where $n \ll p$. These models express the observed data as a linear combination of latent factors where their number, $k$, is much lower than the number of variables $p$ \citep{DavidBartholomewMartinKnott2011}. In the context of metabolomics data, factor analysis models have been used, for example, in molecule classification \citep{Huang2018}, missing data imputation \citep{Shah2019}, biomarker identification \citep{dangelo2021}, and dynamic modelling of the metabolome \citep{Nordin2024}. Here, we propose a novel variant of the infinite factor analysis model \citep{Bhattacharya2011}, designed to model metabolomics data and impute missing values in a statistically and, importantly, physically principled manner. To achieve this, we adopt a truncated factor analysis model congruent with the physical properties of the data.

\subsection{The TGIFA model}
\label{section:model_def}

The general form of a factor model is
\begin{equation}
    \label{equ:model_def}  \boldsymbol{y}_i=\boldsymbol{\mu}+\boldsymbol{\Lambda} \boldsymbol{\eta}_i + \boldsymbol{\epsilon}_i , \quad i=1, \dots, n
\end{equation}
where $\boldsymbol{y}_i = (y_{i1}, \dots, y_{ip})^\top$ is the $p$-dimensional data vector of observation $i$ in the $n \times p$ dataset $\boldsymbol{Y}$, $\boldsymbol{\mu}$ is a $p$-dimensional mean vector, $\boldsymbol{\Lambda}$ is a $p \times k$ loadings matrix, $\boldsymbol{\eta}_i$ is a $k$-dimensional latent factor score for observation $i$, typically assumed to be $N_k(\boldsymbol{0}, \boldsymbol{\operatorname{I}}_k)$, and $\boldsymbol{\epsilon}_i$ is the idiosyncratic error, assumed to be $N_p(\boldsymbol{0}, \boldsymbol{\Sigma})$ where $\boldsymbol{\Sigma} = \operatorname{diag}(\sigma_1^2, \dots, \sigma_p^2)$. The joint distribution of $\left[ \boldsymbol{y}_i, \boldsymbol{\eta}_i \right]^{\top}$ is
\begin{equation*}
\left[\begin{array}{c}
\boldsymbol{y}_i \\
\boldsymbol{\eta}_i
\end{array}\right] \sim
N_{p+k}
    \left(\left[\begin{array}{l}
\boldsymbol{\mu} \\
\boldsymbol{0}
\end{array}\right], \left[\begin{array}{cc}
\boldsymbol{\Lambda} \boldsymbol{\Lambda}^{\top} + \boldsymbol{\Sigma} & \boldsymbol{\Lambda} \\
\boldsymbol{\Lambda}^{\top} & \boldsymbol{\operatorname{I}}_k
\end{array}\right]\right),
\end{equation*}
and under this model, conditional on $\boldsymbol{\eta}_i$, the observed variables $y_{ij}$, $j = 1, \dots, p$, are independent.
Often, $k$ is assumed to be finite and inferred from the data using, for example, an information criterion (e.g., \cite{McNicholas2008}), however, infinite factor models assume $k$ to be potentially infinite and employ shrinkage priors on the loadings matrix \citep{Bhattacharya2011}, obviating the need to fit and choose between multiple models with different values of $k$. Such models have received much attention in the literature, with recent work on e.g., generalised infinite factor models \citep{Schiavon2022}, novel shrinkage techniques \citep{Legramanti2020, Fruhwirth-Schnatter2023} and their use in the context of clustering \citep{murphy2020}.

Metabolomics LC-MS data are strictly non-negative and therefore the typical Gaussian assumption of a general factor model is inappropriate for imputation as it can lead to negative imputed values which are neither physically meaningful nor useful. We therefore introduce a truncated Gaussian infinite factor analysis (TGIFA) model, which assumes that the data, $\boldsymbol{Y}$, are modelled as in (\ref{equ:model_def}) but are jointly distributed with the factor scores according to a truncated multivariate Gaussian distribution. 
Utilising results on partitioning a truncated  multivariate Gaussian distribution \citep{Horrace2005}, we apply truncation below $\boldsymbol{c} = (\boldsymbol{c}_{\boldsymbol{y}}, \boldsymbol{c}_{\boldsymbol{\eta}}) = (\boldsymbol{0}_p, \boldsymbol{-\infty}_k)^\top$ to the joint distribution of  $\left[ \boldsymbol{y}_i, \boldsymbol{\eta}_i \right]^{\top}$, where $\boldsymbol{0}_p$ denotes a vector with $p$ entries of 0 and $\boldsymbol{-\infty}_k$ indicates a vector with $k$ entries of $-\infty$. We denote the resulting truncation of $\left[ \boldsymbol{y}_i, \boldsymbol{\eta}_i \right]^{\top}$ as $\left[ \boldsymbol{y}_i^{t}, \boldsymbol{\eta}_i^{t} \right]^{\top}$. Therefore,
\begin{equation*}
\left[\begin{array}{c}
\boldsymbol{y}_i^{t} \\
\boldsymbol{\eta}_i^{t}
\end{array}\right] \sim
N^{\boldsymbol{c}}_{p+k}
    \left(\left[\begin{array}{l}
\boldsymbol{\mu} \\
\boldsymbol{0}
\end{array}\right], \left[\begin{array}{cc}
\boldsymbol{\Lambda} \boldsymbol{\Lambda}^{\top} + \boldsymbol{\Sigma} & \boldsymbol{\Lambda} \\
\boldsymbol{\Lambda}^{\top} & \boldsymbol{\operatorname{I}}_k
\end{array}\right]\right),
\end{equation*}
where $N^{\boldsymbol{c}}_{p+k}$ indicates a Gaussian distribution of dimension $p+k$ truncated below $\boldsymbol{c}$. Constraints on the distributions of the latent factor scores or idiosyncratic errors, or on the mean vector or loadings are not required. Following \cite{Horrace2005}, we can partition the joint distribution of $\left[ \boldsymbol{y}_i^{t}, \boldsymbol{\eta}_i^{t} \right]^{\top}$  
such that, conditionally,
\begin{equation}
\label{equ:lik}
\begin{array}{rl}
    \boldsymbol{y}_i^{t} \mid  \boldsymbol{\eta}_i^{t} \sim & N_p^{\boldsymbol{c}_{\boldsymbol{y}}}(\boldsymbol{\mu} + \boldsymbol{\Lambda} \boldsymbol{\eta}_i^{t}, \boldsymbol{\Sigma}) \\ 
    = &
    \dfrac{ \exp \big\{ -\frac{1}{2} (\boldsymbol{y}_i^{t} - (\boldsymbol{\mu} + \boldsymbol{\Lambda} \boldsymbol{\eta}_i^{t})) ^{\top} \boldsymbol{\Sigma}^{-1} (\boldsymbol{y}_i^{t} - (\boldsymbol{\mu} + \boldsymbol{\Lambda} \boldsymbol{\eta}_i^{t}))\big\}}{\int^{\boldsymbol{\infty}_p}_{\boldsymbol{c}_{\boldsymbol{y}}} \exp \big\{ -\frac{1}{2} (\boldsymbol{y}_i^{t} - (\boldsymbol{\mu} + \boldsymbol{\Lambda} \boldsymbol{\eta}_i^{t})) ^{\top} \boldsymbol{\Sigma}^{-1} (\boldsymbol{y}_i^{t} - (\boldsymbol{\mu} + \boldsymbol{\Lambda} \boldsymbol{\eta}_i^{t}))\big\} d \boldsymbol{y}_i^{t}}. \\
    \end{array}
\end{equation}
Thus the desirable property that $y_{ij}^{t}$ and $y_{ij'}^{t}$ are conditionally independent given $\boldsymbol{\eta}_i^{t}$ for $j \neq j'$ and $j = 1, \dots, p$, which arises naturally in the unconstrained factor analysis model \citep{Robert1995, Kotecha1999, Rodriguez-Yam2004}, holds for TGIFA .

The marginal distribution of $\boldsymbol{\eta}_i^{t}$ is not available in an easily tractable form, as marginal distributions resulting from partitioning a truncated multivariate Gaussian distribution are not truncated Gaussian except in specific circumstances \citep{Horrace2005b}. Specifically, 
\begin{equation}
\label{equ:eta_marginal}
    \boldsymbol{\eta}_i^{t} \sim 
    \frac{ \int^{\boldsymbol{\infty}_p}_{\boldsymbol{c}_{\boldsymbol{y}}} \exp \big\{ -\frac{1}{2} [\boldsymbol{y}_i^{t} - \boldsymbol{\mu}, \boldsymbol{\eta}_i^{t}]  \boldsymbol{\Sigma}^{*-1} [\boldsymbol{y}_i^{t} - \boldsymbol{\mu}, \boldsymbol{\eta}_i^{t}]^{\top}\big\} d \boldsymbol{y}_i^{t}}    {\int^{\boldsymbol{\infty}_{p+k}}_{\boldsymbol{c}} \exp \big\{ -\frac{1}{2} [\boldsymbol{y}_i^{t} - \boldsymbol{\mu}, \boldsymbol{\eta}_i^{t}]  \boldsymbol{\Sigma}^{*-1} [\boldsymbol{y}_i^{t} - \boldsymbol{\mu}, \boldsymbol{\eta}_i^{t}]^{\top} \big\} d [\boldsymbol{y}_i^{t}, \boldsymbol{\eta}_i^{t}]^{\top}},
\end{equation}
where $\boldsymbol{\Sigma}^{*-1}$ denotes the precision matrix of the joint distribution of $[\boldsymbol{y}_i, \boldsymbol{\eta}_i]^\top$. Although it follows that the marginal distribution of $\boldsymbol{y}_i^{t}$ is also not easily tractable, it is not required under the inferential approach detailed in Section \ref{section:inf}.

\subsection{Accounting for missing data}

To jointly model MAR and MNAR missing data mechanisms, we introduce a missingness indicator for observation $i$ and variable $j$, $r_{ij}$, with $r_{ij}=0$ for missing $y_{ij}$, and $r_{ij}=1$ for observed $y_{ij}$. As summarised in Table \ref{tab:r_def}, given the LOD, if $y_{ij} < \text{LOD}$, $r_{ij} = 0$ with probability $1$ and $r_{ij} = 1$ with probability $0$, whereas if $y_{ij} > \text{LOD}$, then $r_{ij} = 0$ with probability $\alpha$ and $r_{ij} = 1$ with probability $1 - \alpha$.

\begin{table}[htbp]
\centering
\caption{Probability mass function of the missingness indicator $r_{ij}$ given $y_{ij}$ and the $\text{LOD}$.}
\begin{tabular}{|c||c|c|}
\hline
  & $y_{ij} < \text{LOD}$ & $y_{ij} > \text{LOD}$ \\ \hline \hline
$r_{ij} = 0$ & 1 & $\alpha$     \\ \hline
$r_{ij} = 1$ & 0 & $1 - \alpha$ \\ \hline
\end{tabular}
\label{tab:r_def}
\end{table}

Let $\boldsymbol{\theta} = \big\{\boldsymbol{\mu}, \{\boldsymbol{\lambda}_j\},
\{\sigma_j^{-2}\}, 
\alpha\big\}$ denote model parameters, where $\boldsymbol{\lambda}_j = \{\lambda_{j1}, \dots, \lambda_{jk}\}^{\top}$, is a row of the loadings matrix. Considering $\boldsymbol{R} = \{\boldsymbol{r}_1, \dots, \boldsymbol{r}_n\}^\top$, the $n \times p$ observed matrix where $\boldsymbol{r}_i = (r_{i1}, \dots, r_{ip})$, the joint distribution of $\boldsymbol{Y}$ and $\boldsymbol{R}$  conditional on $\boldsymbol{\theta}$ and $\boldsymbol{\eta} = \{\boldsymbol{\eta}_1, \dots, \boldsymbol{\eta}_n \}$ is
\begin{equation}
\label{equ:y_r_joint}
\begin{array}{rrl}
     \operatorname{p}(\boldsymbol{Y}, \boldsymbol{R} \mid \boldsymbol{\theta}, \boldsymbol{\eta}) = 
      & \displaystyle \prod_{i=1}^n \prod_{j=1}^p \Big[ & \big[ (1 - \alpha ) \operatorname{p}(y_{ij} \mid y_{ij} > \text{LOD}, \boldsymbol{\theta}, \boldsymbol{\eta}_i) \big]^{\mathds{1}\{r_{ij}=1\}} \\
     & & \big[ \operatorname{p}(y_{ij} \mid y_{ij} < \text{LOD}, \boldsymbol{\theta}, \boldsymbol{\eta}_i) \\
     & & + \alpha \operatorname{p}(y_{ij} \mid y_{ij} > \text{LOD}, \boldsymbol{\theta}, \boldsymbol{\eta}_i) \big]^{\mathds{1}\{r_{ij}=0\}} \Big],
\end{array}
\end{equation}
where $\mathds{1}\{\cdot\}$ is an indicator function.

\subsection{Prior distributions}
\label{section:mod_priors}

Inference proceeds via a Bayesian framework. The following prior distributions are assumed, where $\operatorname{Ga}(\alpha, \beta)$ refers to the gamma distribution whose mean is given by ${\alpha/\beta}$:
\begin{eqnarray*}
 \boldsymbol{\mu} & \sim &  N_p(\boldsymbol{\tilde{\mu}}, \boldsymbol{\varphi}^{-1} \boldsymbol{\operatorname{I}}_p),\\
\lambda_{jh} & \sim & N_1 (0, \phi_{jh}^{-1}\tau_h^{-1}) \:\: \mbox{ for $j = 1, \dots, p$ and $h = 1, \dots, \infty$},\\
\sigma_{j}^{-2} & \sim & \operatorname{Ga}(a_{\sigma}, b_{\sigma}) \:\: \mbox{ for $j = 1, \dots, p$}, \\
\alpha & \sim & \operatorname{Unif}(0, 1).
\end{eqnarray*}
To facilitate shrinkage of the factor loadings towards zero as the factor dimension $h$ increases, a multiplicative truncated gamma process shrinkage prior (MTGP) \citep{Gwee2025}, is assumed for the variance of the prior on $\lambda_{jh}$. Specifically, a prior is assumed on $\phi_{jh}$, which acts as a local shrinkage parameter, and on $\tau_h$, which acts as a column-wise shrinkage parameter, i.e., 
\begin{eqnarray*}
    \phi_{j h} \sim \operatorname{Ga}(\kappa_1, \kappa_2),
    \quad 
     \tau_{h}=\prod_{l=1}^{h} \delta_{l},\\
     \delta_{1} \sim \operatorname{Ga}\left(a_{1}, 1\right), \quad \delta_{l} \sim \operatorname{Ga}^{[1, \infty)}\left(a_2, 1\right)\quad l = 2, \dots, \infty.
\end{eqnarray*}
The nature of the MTGP which places a gamma distribution prior, truncated to values above 1, on $\delta_l$ for dimensions $l = 2, \dots, \infty$, ensures the desired shrinkage behaviour by constraining these $\delta_l$ to values of $1$ or greater. Care must also be taken when specifying values for hyperparameters $a_1$ and $a_2$, as discussed in \cite{Durante2017}. The utility of assuming $k = \infty$ and applying such a shrinkage prior is that the need to fit multiple models with different numbers of latent factors, and to select and use model selection criteria to choose the optimal model, is obviated.

\section{Inference and imputation}
\label{section:inf}

\subsection{Inference for the truncated multivariate Gaussian distribution via the exchange algorithm}

A Bayesian approach to inference and imputation for the TGIFA model proceeds by employing a Markov chain Monte Carlo (MCMC) sampler to explore the posterior distribution. Given the likelihood function and prior distributions, the posterior distribution is 
\begin{equation*}
\begin{split}
    \operatorname{p}( \boldsymbol{\mu}, \boldsymbol{\Lambda}, \boldsymbol{\Sigma}, \alpha, \boldsymbol{\eta}, \boldsymbol{\phi}, \boldsymbol{\delta} \mid \boldsymbol{Y}, \boldsymbol{R}) \propto & \operatorname{p}(\boldsymbol{Y}, \boldsymbol{R} \mid \boldsymbol{\mu}, \boldsymbol{\Lambda}, \boldsymbol{\Sigma}, \alpha, \boldsymbol{\eta}) \operatorname{p}(\boldsymbol{\mu}) \operatorname{p}(\boldsymbol{\Lambda} \mid \boldsymbol{\phi}, \boldsymbol{\delta}) \operatorname{p}(\boldsymbol{\Sigma})  \operatorname{p}(\alpha) \\ 
    &   \operatorname{p}(\boldsymbol{\eta}) \operatorname{p} (\boldsymbol{\phi})\operatorname{p}(\boldsymbol{\delta}),
    \end{split}
\end{equation*}
where 
$\boldsymbol{\phi} = \{\boldsymbol{\phi}_1, \dots, \boldsymbol{\phi}_{p}\}$ with $\boldsymbol{\phi}_j = (\phi_{j1}, \dots, \phi_{jk})^\top$ for $j = 1, \dots, p$, 
and $\boldsymbol{\delta} = (\delta_{1}, \dots, \delta_{k})^\top$.

The Gibbs sampling inferential procedure typically used for a standard Gaussian infinite factor model is not available for TGIFA as inference is complex, largely due to the high-dimensional integrals in the conditional and marginal distributions of the truncated Gaussian distribution. Therefore, here we propose the use of the exchange algorithm \citep{murray2006} for inference for the TGIFA model. The exchange algorithm is often used in the case of doubly intractable posterior distributions, where a standard Metropolis-Hastings (MH) algorithm cannot be used. In such cases, the model likelihood is $p(\boldsymbol{Y} | \boldsymbol{\theta}) = {\tilde{p}(\boldsymbol{Y} | \boldsymbol{\theta})}/{\mathcal{Z}(\boldsymbol{\theta})}$, say, where $\mathcal{Z}(\boldsymbol{\theta})$ is intractable, along with the usual intractable normalisation constant of the posterior distribution. The exchange algorithm neatly obviates the need to compute $\mathcal{Z}(\boldsymbol{\theta})$ by sampling an auxiliary observation $\breve{\boldsymbol{Y}}$ from $p(\boldsymbol{Y}|\boldsymbol{\theta})$ and incorporating it in the MH algorithm's acceptance probability,  which results in cancellation of the intractable $\mathcal{Z}(\boldsymbol{\theta})$ terms.
While the exchange algorithm has been extensively used in a range of contexts involving doubly intractable posterior distributions, for example in Gaussian graphical models \citep{Mohammadi2015} and in rank data modelling \citep{Piancastelli2024}, here we employ it in the context of truncated Gaussian distributions to enable scalable inference in high-dimensional truncated models. 

During inference for TGIFA, evaluation of the truncated Gaussian density $p(\boldsymbol{y}_i^t \mid \boldsymbol{\eta}_i^t)$ from (\ref{equ:lik}), for example, is required; while tractable, it is computationally expensive to compute due to the dimensionality of the denominator's integral. The exchange algorithm is therefore employed to improve computational performance by overcoming the need to evaluate this integral. Specifically, an auxiliary observation $\breve{\boldsymbol{Y}}^t$ is easily simulated from the truncated Gaussian in (\ref{equ:lik}), allowing cancellation of its denominator in the MH acceptance probability.  While full details of the inferential procedure are available in Appendix \ref{app:FIP}, an outline is provided in Algorithm \ref{alg:exchange} where $\operatorname{p}(\boldsymbol{\theta} \mid \dots)$ denotes the conditional posterior distribution of $\boldsymbol{\theta}$ given all other model parameters, and $\operatorname{q}(\boldsymbol{\theta} \mid \dots)$ denotes the respective proposal distributions. In summary, the factor scores are updated using MH, the mean, factor loadings, and variance of the idiosyncratic errors are updated using the exchange algorithm, and the shrinkage and missingness parameters are updated using a Gibbs sampler. As inference on the number of latent factors is not of interest here, $k^{*}$ is used throughout as a finite, conservatively large number of latent factors. We therefore exploit the exchange algorithm as a novel, computationally-efficient approach to inference with truncated multivariate Gaussian distributions. 

\subsection{Imputation of missing data}

Missing values, denoted $\ddot{y}_{ij}$, are imputed at each MCMC iteration. Naturally, only entries $\ddot{y}_{ij}$ where $r_{ij}=0$ are imputed, such that 

\begin{equation*}
    \begin{array}{rl}
        \operatorname{p}(r_{ij} = 0, \ddot{y}_{ij}) = 
      & \operatorname{p}(\ddot{y}_{ij} < \text{LOD}) + \alpha \operatorname{p}(\ddot{y}_{ij} > \text{LOD}) \\
      = & \int_{0}^{\text{LOD}} N_1 ^{[0, \infty)} \left(\ddot{\mu}_{ij}, \ddot{\sigma}_{ij}^2\right)d\ddot{y}_{ij} + \alpha \int^{\infty}_{\text{LOD}} N_1 ^{[0, \infty)} \left(\ddot{\mu}_{ij}, \ddot{\sigma}_{ij}^2\right)d\ddot{y}_{ij} \\
        = & P + \alpha Q,
    \end{array}
\end{equation*}
where
$\ddot{\mu}_{ij}$ denotes the $ij^{th}$ element of the current value of $\boldsymbol{\mu} + \boldsymbol{\Lambda} \boldsymbol{\eta}_i$, and $\ddot{\sigma}_{ij}^2$ denotes the current value of the $j^{th}$ diagonal element of $\boldsymbol{\Sigma}$. Given that $r_{ij}=0$, a binary latent variable $z_{ij}$, is then defined such that $z_{ij} = 0$ if $\ddot{y}_{ij} > \text{LOD}$ (i.e., MAR), and $z_{ij} = 1$ if $\ddot{y}_{ij} < \text{LOD}$ (i.e., MNAR). By Bayes' theorem then
\begin{equation*}
    z_{ij} \mid r_{ij} = 0, \dots \sim \operatorname{Bernoulli} \Big[\frac{P}{P + \alpha Q} \Big]
\end{equation*}
and $\ddot{y}_{ij} \mid z_{ij}= 1, \dots  \sim N_1 ^{[0, \text{LOD})} (\ddot{\mu}_{ij}, \ddot{\sigma}_j^2)$ and $\ddot{y}_{ij} \mid z_{ij} = 0, \dots \sim N_1 ^{[\text{LOD}, \infty)} (\ddot{\mu}_{ij}, \ddot{\sigma}_j^2)$. Missing values in $\boldsymbol{Y}$ are therefore imputed independently at each iteration of the MCMC chain, by sampling $z_{ij}$ and then $\ddot{y}_{ij} \mid z_{ij}, \dots$, giving an updated complete dataset. Once the MCMC chain has converged, after discarding burn-in iterations, the posterior modal missingness designation (MAR or MNAR) is available for each missing value, and the posterior median of the imputed values under that modal designation is used as the final imputed value. Importantly, the Bayesian approach naturally provides credible intervals, quantifying the uncertainty in both the imputed value and the missingness designation.

\begin{algorithm}
\caption{The TGIFA inferential procedure and imputation. Further detail on this algorithm is available in Appendix \ref{app:FIP}.}
\label{alg:exchange}
    \begin{algorithmic}
    \Require Initial values of all model parameters, number of MCMC iterations, \texttt{n.iters}, and proposal distributions $\operatorname{q}(\breve{\theta} \mid \theta)$ for $\theta \in \boldsymbol{\theta}_{-\alpha}$.
        \For{$m = 1$ to \texttt{n.iters}}
        \For{$\theta$ in $\boldsymbol{\theta}_{-\alpha}$}
            \State Sample a proposal value $\breve{\theta} \sim \operatorname{q}(\breve{\theta} \mid \theta)$.
            \State Sample an auxiliary observation $\breve{\boldsymbol{Y}}^t \sim \operatorname{p}(\boldsymbol{Y}^{t} \mid \breve{\boldsymbol{\theta}}, \boldsymbol{\eta})$.
            \State Compute acceptance probability, $\mathcal{A}_{\theta}$.
            \State Accept proposed $\breve{\theta}$ with probability $\mathcal{A}_{\theta}$ to set $\theta = \breve{\theta}$.
         \EndFor
         \State Update $\boldsymbol{\eta}$ via MH.
         \State Update $\boldsymbol{\phi}, \boldsymbol{\delta}, \alpha$ via Gibbs sampling.
         \State Update imputed values.
         \EndFor
    \end{algorithmic}
\end{algorithm}

\section{Simulation study}
\label{section:simulation}

A simulation study was conducted to assess the imputation performance of the TGIFA approach and to compare its performance to state-of-the-art imputation methods.

\subsection{Simulation study set up}
\label{section:simulation_set_up}
To assess the performance of TGIFA, we conducted a simulation study comparing its imputation accuracy and uncertainty quantification capabilities to several benchmark methods.
The simulations were designed to mimic key characteristics of the urinary metabolomics dataset (see Section \ref{section:broccoli}), including high dimensionality ($n \ll p$), multicollinearity, non-negativity, and a combination of MAR and MNAR missingness mechanisms. We generated ten datasets, each with $n = 18$ and $p = 1391$, from the TGIFA model (\ref{equ:lik}). For each dataset, principal component analysis (PCA) was applied to the fully-observed variables of the urinary metabolomics dataset and the $\boldsymbol{\Lambda}$ used to generate the dataset was fixed as the first $k^* = 5$ components of the resulting loadings matrix, which explained 70\% of the data's total variance. We attributed $2(100 - 70)=60\%$ of the simulated data's variance to the idiosyncratic errors to emulate the true variance of the real data, setting $\sigma^{2}_{j}$ to $0.6$ of the variance of the $j^{\text{th}}$ variable of the urinary metabolomics dataset.
For observation $i$, $\boldsymbol{\eta}_i \sim N_{k^{*}}(\boldsymbol{0}, \textbf{I}_{k^{*}})$.
The mean $\boldsymbol{\mu}$ was simulated from the prior specified in Section \ref{section:mod_priors}. The $\boldsymbol{\tilde{\mu}}$ hyperparameter was set by subtracting the row mean of $\boldsymbol{\Lambda} \boldsymbol{\eta}^{\top}$ from the sample mean of the fully-observed variables from the urinary metabolomics dataset and $\boldsymbol{\varphi}^{-1}$ was set to a value of $0.05$ times the same sample mean.

Missingness under the two missingness mechanisms was then introduced to emulate the total missingness proportion in the urinary metabolomics dataset. For MNAR missingness, all values below the 0.015 quantile of each simulated dataset were replaced by $\texttt{NA}$, such that $1.5\%$ of each dataset was MNAR.
For MAR missingness, in each simulated dataset, 1.5\% of the remaining non-missing entries were randomly selected and replaced by $\texttt{NA}$ values resulting in a total missingness proportion of $3\%$ per dataset.

The TGIFA model was fitted to each simulated dataset using R \citep{RCoreTeam2024}, running MCMC chains for $10000$ iterations, with a $5000$ burn-in and thinned to every $5^{\text{th}}$ iteration.  The hyperparameters were set following \cite{Bhattacharya2011} and \cite{murphy2020}: $\kappa_1 = 3$, $\kappa_2=2$, $a_{\sigma}=1$, $b_{\sigma}=0.25$, and, for the MTGP, $a_1=2.1$, $a_2=3.1$, as in \cite{Durante2017} and \cite{Gwee2025}. Imputed values were initialised as the absolute values of SVD imputed values using the \texttt{softImpute} R package \citep{TrevorHastie2021}. The number of latent factors $k^{*}=5$ during model fitting, following the logic used in data simulation. While use of an adaptive Gibbs sampler \citep{Bhattacharya2011}, which facilitates the addition and removal of factors as the MCMC evolves, was explored, little inferential benefit and greater computational cost was observed. Initial values of $\boldsymbol{\Lambda}$ were the simulated data set's PCA loadings and $\boldsymbol{\Sigma}$ and $\boldsymbol{\eta}$ were initialised using the same approach as in the data simulation. Finally, $\boldsymbol{\mu}$ was initialised as the sample mean with the row means of $\boldsymbol{\Lambda} \boldsymbol{\eta}^{\top}$ subtracted, with $\boldsymbol{\tilde{\mu}} = \boldsymbol{\mu} - 1$ to achieve acceptable acceptance probabilities for $\boldsymbol{\mu}$.
For  $j = 1, \dots, p$, 
$\varphi^{-1}_j$ was set to 0.05 times the sample mean of variable $j$, where variable $j$ contained missing values, and $\varphi^{-1}_j = 1$ where variable $j$ was fully observed.
The remaining parameters were initialised from their respective priors.

For comparative purposes, missing values were also imputed using fixed-value imputation methods (half-minimum and mean imputation), SVD imputation, RF imputation using the \texttt{missForest} R package \citep{Stekhoven2022}, and infinite factor analysis (IFA) imputation (similar to the approach proposed in \cite{Shah2019}) applied to both the original and log-transformed data (denoted here as $\text{IFA}$ and $\text{logIFA}$  respectively). Where relevant, initial values and hyperparameter settings in the IFA methods were set in the same manner as the TGIFA method. To assess performance, for each simulated dataset the mean absolute error (MAE) and residuals between posterior median imputed values and true values were computed for all methods considered. Where available, credible intervals are also reported. 

\subsection{Simulation study results}

\begin{figure}[htp]
\begin{subfigure}{\textwidth}
  \centering
  \includegraphics[height=0.2\textheight, keepaspectratio]{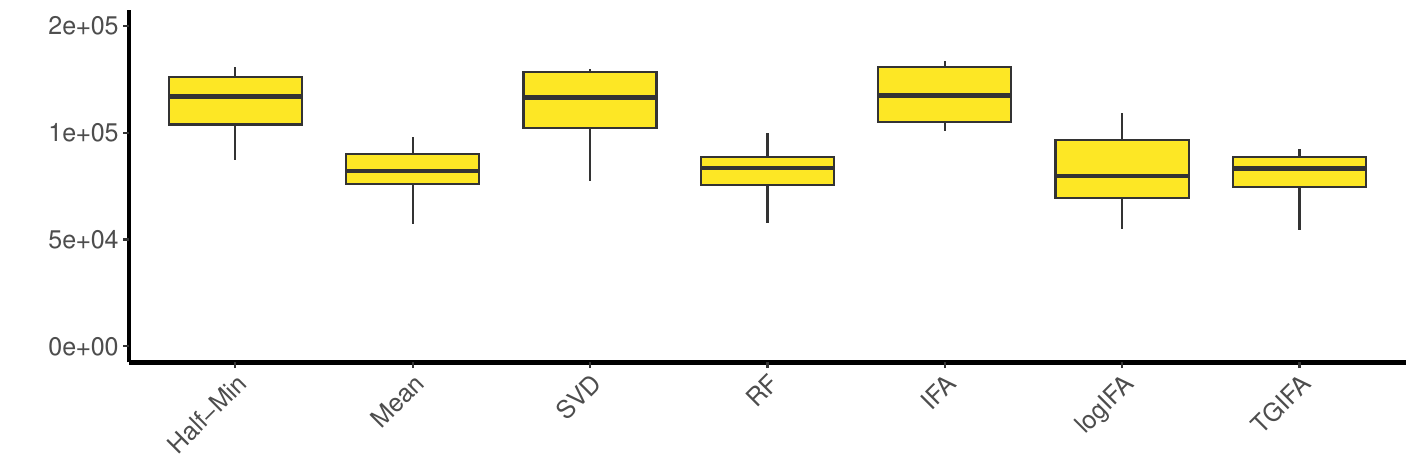}
  \caption{}
  \label{fig:mae_boxplots_all}
\end{subfigure}
\begin{subfigure}{\textwidth}
  \centering\includegraphics[height=0.2\textheight, keepaspectratio]{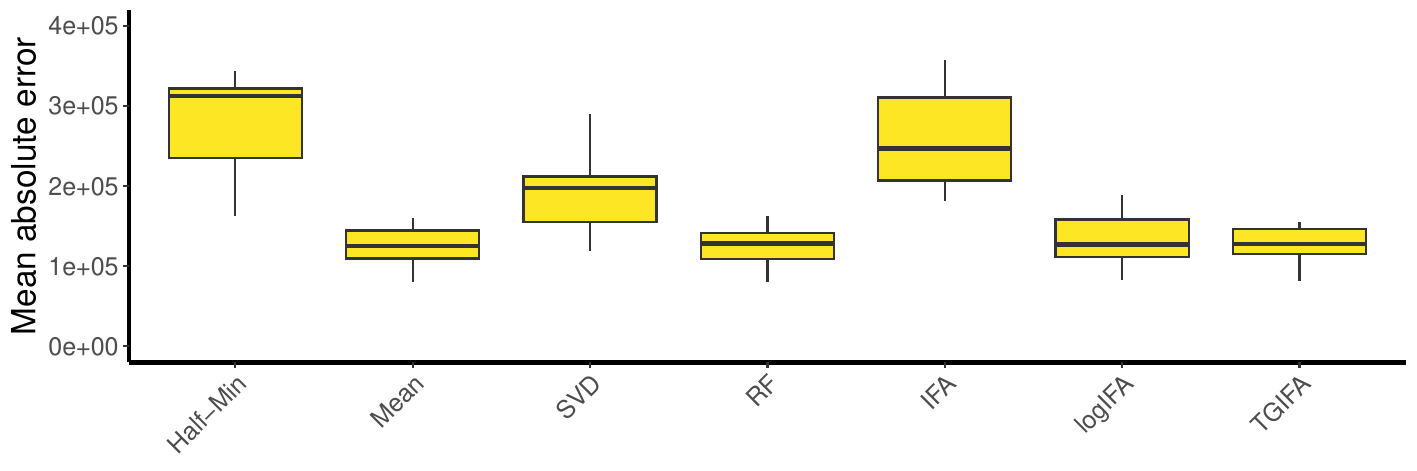}
  \caption{}
  \label{fig:mae_boxplots_MAR}
\end{subfigure}
\begin{subfigure}{\textwidth}
  \centering
\includegraphics[height=0.2\textheight, keepaspectratio]{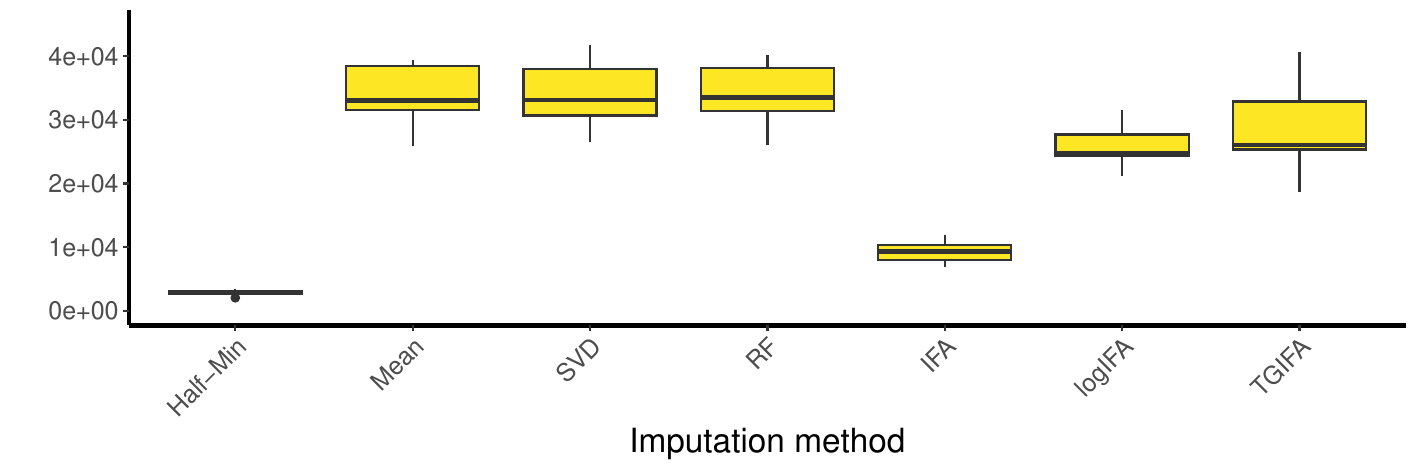}
  \caption{}
    \label{fig:mae_boxplots_MNAR}
\end{subfigure}
\caption{Mean absolute errors between posterior
median imputed values and true values across ten simulated datasets for (a) all imputed values, (b) MAR imputed values and (c) MNAR imputed values, across all imputation methods. 
}
\label{fig:mae_boxplots}
\end{figure}

Of the fixed-value imputation methods, the half-minimum approach exhibits poorer performance than mean imputation across all imputed values (see Figure \ref{fig:mae_boxplots_all}), with larger and more variable MAEs. While the RF method performs similarly to mean imputation, the SVD and IFA approaches perform poorly, with MAEs larger and more variable than RF imputation. The logIFA approach performs well on average, however, exhibits large variability in MAE, and failed to complete for one of the ten simulation replicates due to ill-conditioned covariance matrix estimates caused by rounding errors. The TGIFA method performs similarly to mean and RF imputation, with similar MAE values and variability. A similar performance pattern is observed when imputation of MAR entries only is considered (Figure \ref{fig:mae_boxplots_MAR}), where the MAR designation refers to the true missing data mechanism. Of note is that the IFA method performs particularly poorly, as true MAR entries are designated as MNAR with greater frequency than under the logIFA and TGIFA models. 

Imputation performance on the MNAR entries, where the MNAR designation refers to the true missing data mechanism, exhibits a different trend (Figure \ref{fig:mae_boxplots_MNAR}). The half-minimum imputation method performs well, as this method imputes fixed-values which, due to the LOD in LC-MS data, will not be far from the truth. The mean, SVD, and RF methods exhibit similar results, with higher MAEs than the fixed-value methods. The IFA method appears to perform well, however, the low MAEs come at a cost as physically impossible negative values are imputed. As the IFA model allows such values, we do not truncate negative values to zero, or to exclude them, when calculating the MAE. The same is true of SVD imputation, without the performance boost. The logIFA and TGIFA methods show similar MNAR imputation performance to the mean and RF methods, with the lowest median MAEs. Naturally, for all methods, the magnitude of the MAE values for MNAR imputation is lower than for the MAR case, as the missing values are present in variables with lower measurement values in general.

Summaries of residuals between the posterior median imputed values and true values for a single simulated dataset are available in Appendix \ref{app:simstudy}, along with plots of imputed versus true values for each imputation method. Performance trends in the residuals are similar to those exhibited by the MAE. Posterior mean values of the loadings matrix are also provided for a selection of simulation replicates. 

\begin{figure}[htp]
\centering
\begin{subfigure}{\textwidth}
  \centering
  \includegraphics[height=0.18\textheight, keepaspectratio]{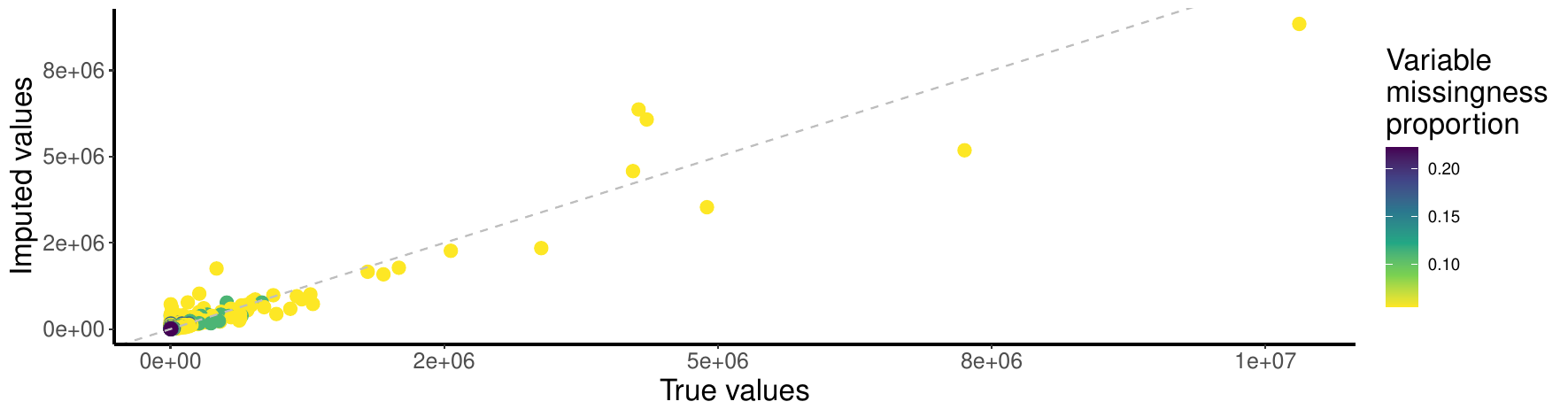}
  \caption{}
  \label{fig:imp_comp_mean}
\end{subfigure}
\begin{subfigure}{\textwidth}
  \centering
  \includegraphics[height=0.18\textheight, keepaspectratio]{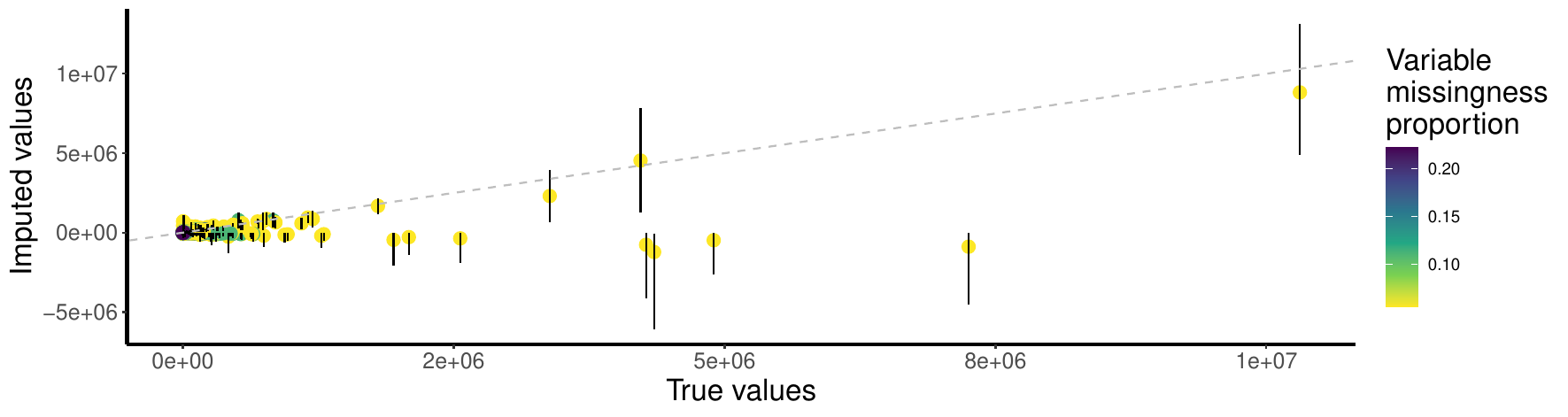}
  \caption{}
  \label{fig:imp_comp_ifa}
\end{subfigure}
\begin{subfigure}{\textwidth}
  \centering
  \includegraphics[height=0.18\textheight, keepaspectratio]{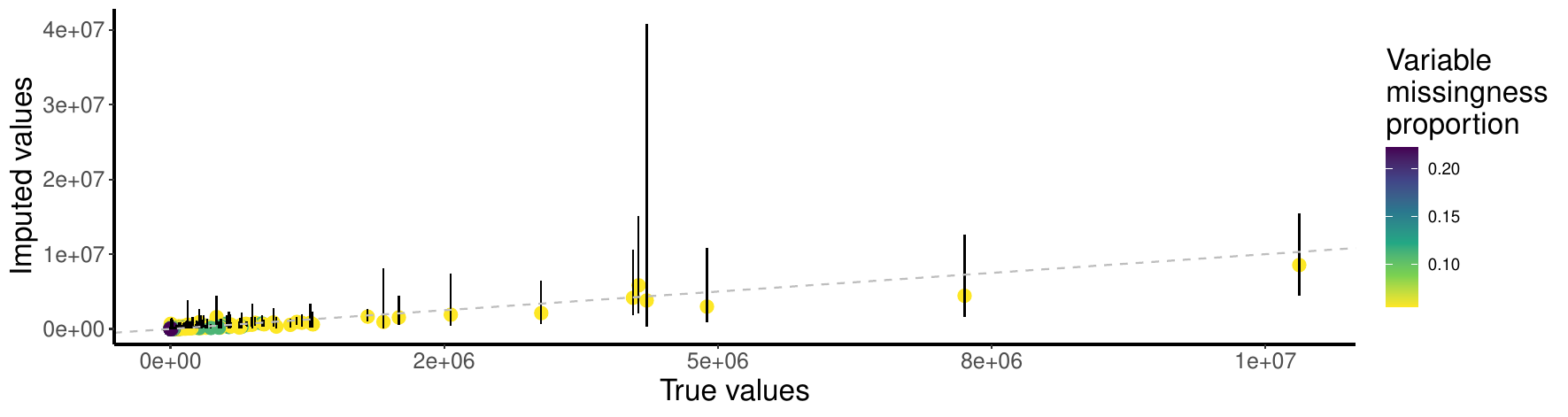}
  \caption{}
  \label{fig:imp_comp_logifa}
\end{subfigure}
\begin{subfigure}{\textwidth}
  \centering
\includegraphics[height=0.18\textheight, keepaspectratio]{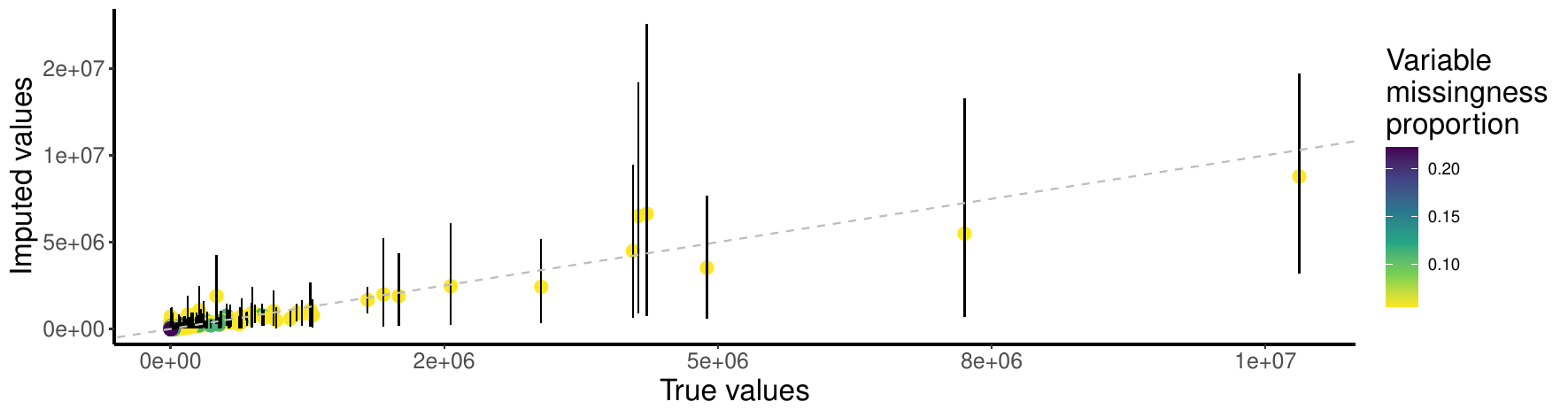}
\caption{}
  \label{fig:imp_comp_tgifa}
  \end{subfigure}
\caption{For one simulated data set, true values versus (a) imputed values under mean imputation and versus posterior median imputed values, and associated $95\%$ credible intervals, under (b) IFA, (c) logIFA, and (d) TGIFA imputation. The dashed grey line is the line of equality and points are coloured by their variable's proportion of missingness. Note that the scale of the y-axis varies between panels.}
\label{fig:imp_comp_ifa_mean_tgifa}
\end{figure}

Comparing true versus imputed values for a single simulated dataset, using the mean, IFA, logIFA, and TGIFA approaches (Figure \ref{fig:imp_comp_ifa_mean_tgifa}), under mean imputation (Figure \ref{fig:imp_comp_mean}), only a point estimate is imputed for each missing value and no measure of uncertainty is readily available; if there are multiple missing entries in one variable the same fixed value is imputed for all of them.
Additionally, imputation below the assumed LOD is not possible under mean imputation meaning that the truly MNAR entries are not appropriately imputed. 
In Figure \ref{fig:imp_comp_ifa}, the IFA approach imputes missing data with impractical negative values, but does quantify the uncertainty associated with the imputed values. Imputation under logIFA, shown in Figure \ref{fig:imp_comp_logifa} shows sensible imputation, but very large credible intervals, as imputation of logged values and subsequent exponentiation amplifies small changes. Another contributing cause is that the lognormal assumptions implicit in logIFA skew the data's distribution towards larger values; this does not tend to skew the posterior median imputed value for each missing value, however, can skew credible interval bounds towards more extreme values.
Using TGIFA, meaningful posterior median imputed values and their associated $95\%$ quantile-based credible intervals provide richer inference for the user (Figure \ref{fig:imp_comp_tgifa}). Similar figures for the other imputation methods considered are provided in Appendix \ref{app:simstudy}.

Table \ref{tab:acc} outlines the overall, MAR, and MNAR missingness designation accuracy for TGIFA and for the comparable methods of IFA and logIFA. The mean overall missingness designation accuracy across all simulated datasets under TGIFA was $74.8\%$, which was well-balanced between MAR and MNAR designation accuracies. In the case of logIFA, the mean overall designation accuracy of 66.8\% was  skewed by strong MAR designation accuracy while in the case of IFA, designation accuracy performance was skewed by strong accuracy in MNAR designations. Overestimating the proportion of MAR entries is generally less problematic, as truly MNAR entries tend to be present in variables with lower values, so the results of an incorrect MAR imputation are generally not as far from the true value than if a truly MAR entry in a larger variable were to be imputed incorrectly as MNAR.

\begin{table}[ht]
\centering
\caption{Missingness type designation accuracy for IFA, logIFA, and TGIFA imputation, across all simulated datasets.}
\begin{tabular}{|c||c|c|c|}
\hline
  & \textbf{Overall} & \textbf{MAR} & \textbf{MNAR} \\ \hline\hline
\textbf{IFA} & $61.3\pm1.2$ & $34.4\pm2.7$ & $90.9\pm1.1$ \\ \hline
\textbf{logIFA} & $66.8\pm1.3$ & $93.9\pm1.6$ & $37.1\pm3.1$ \\ \hline
\textbf{TGIFA} & $74.8\pm0.7$ & $77.2\pm1.6$ & $72.2\pm1.9$ \\ \hline
\multicolumn{4}{l}{\footnotesize *Values represent mean percentage accuracy $\pm$ sd }\\ 
\multicolumn{4}{l}{\footnotesize of percentage accuracy across all simulations.
} \\
\end{tabular}
\label{tab:acc}
\end{table}

\section{Imputing missing values in urinary metabolomics data}
\label{section:real}

We applied the TGIFA model, with $k^* = 5$, to impute missing values in the motivating urinary metabolomics dataset introduced in Section \ref{section:broccoli}. For comparative purposes, imputation via half-minimum, mean, SVD, RF, IFA, and logIFA imputation were also considered. 

\begin{figure}[htp]
\begin{subfigure}{\textwidth}
  \centering
  \includegraphics[width=\textwidth, keepaspectratio]{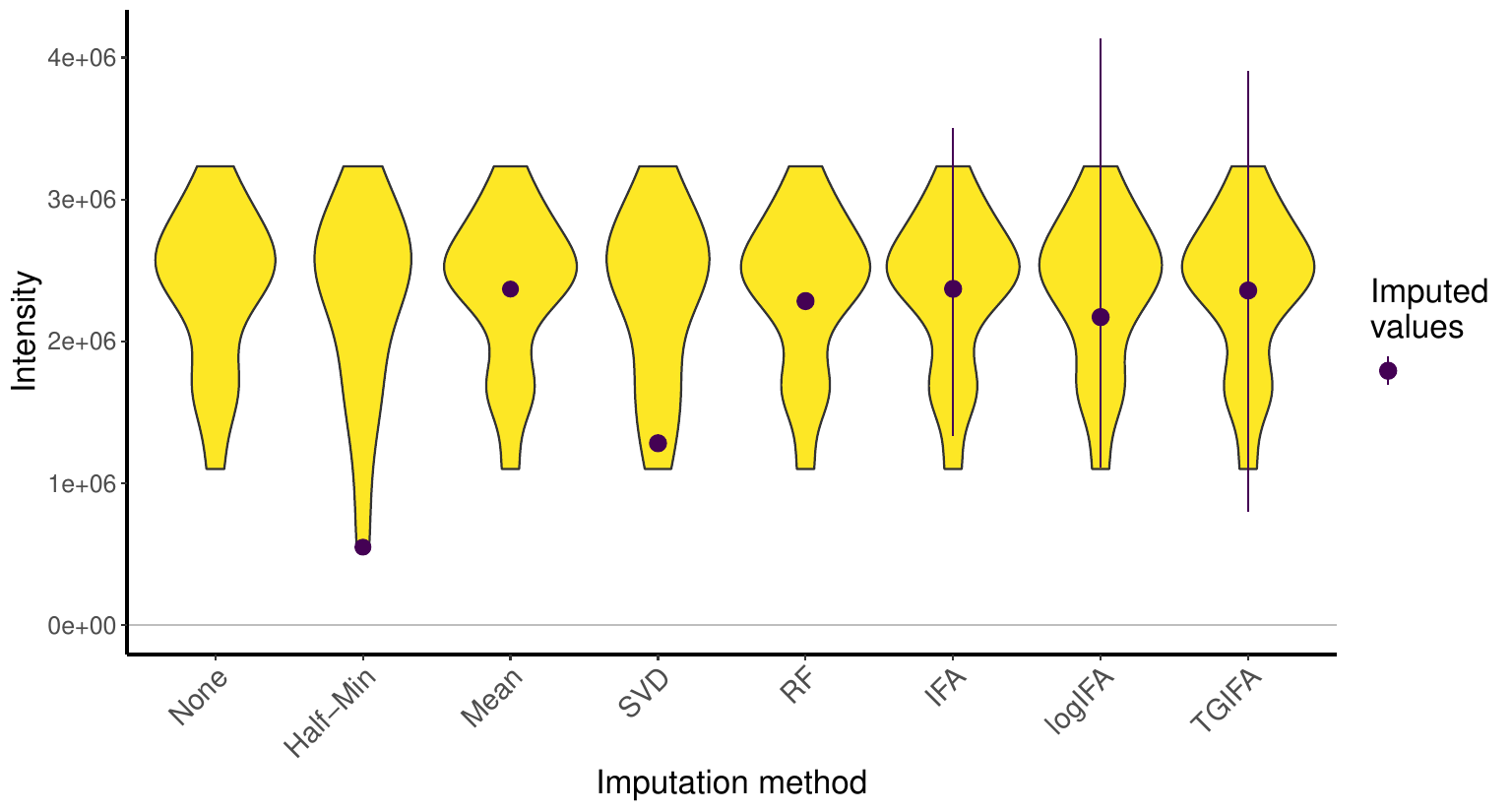}
  \caption{}
\label{fig:broc_1924_violin}
\end{subfigure}
\begin{subfigure}{\textwidth}
  \centering
  \includegraphics[width=\textwidth, keepaspectratio]{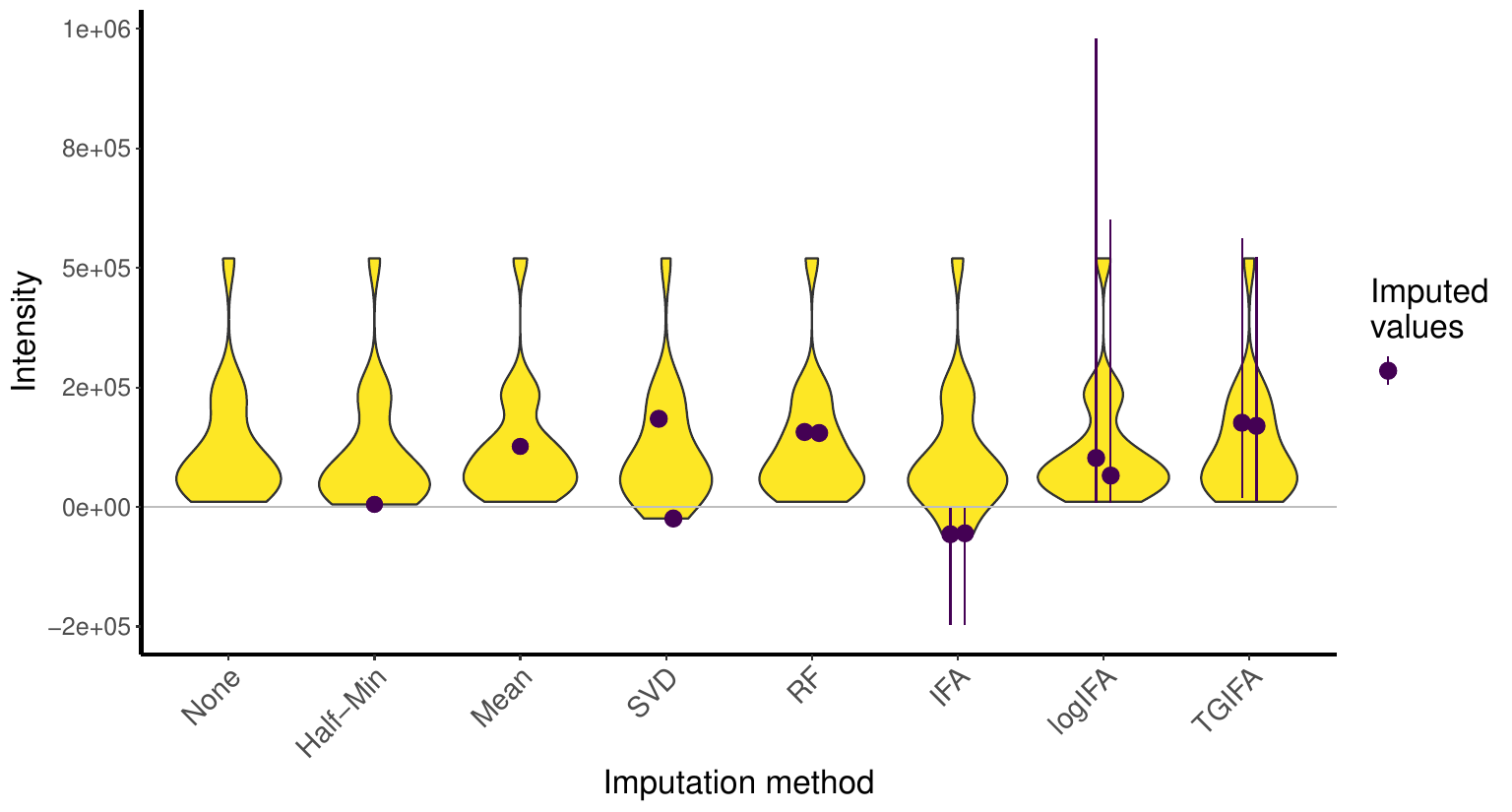}
  \caption{}
\label{fig:broc_1318_violin}
\end{subfigure}
\caption{Violin plots and imputed values for two variables with high mean values and variances from the urinary metabolomics dataset: (a) variable 1924 with one missing value and (b) variable 1318 with two missing values. For half-minimum and mean imputation, only one imputed point is visible, as all missing entries are imputed with the same value. For imputation methods that impute different values for different missing values, imputed values are jittered for clarity. Where available, 95\% credible intervals are provided.}
\label{fig:violin_1924_1318}
\end{figure}

Figure \ref{fig:violin_1924_1318} provides violin plots of two variables that have relatively large values and variances; for such variables it is likely that any missing values would be MAR as the observed values are far from the LOD. For both variables, the fixed-value imputation methods (half-minimum, and mean imputation) do not provide uncertainty quantification, and in the case of variable 1318 they impute the same value for each missing entry. The SVD and IFA approaches impute reasonable values for variable 1924, however, in variable 1318 inappropriate negative values are imputed; the IFA method does, however, provide uncertainty quantification. For the RF approach, plausible values are imputed, but no uncertainty is provided. The logIFA approach imputes feasible values, however, in general, the upper bound of the quantile-based credible intervals can be relatively large, as values far outside the observed range are imputed.
In contrast, the TGIFA model provides physically plausible posterior median imputed values, and $95\%$ credible intervals for both variables.

\begin{figure}[htp]
\begin{subfigure}{\textwidth}
  \centering
  \includegraphics[width=\textwidth, keepaspectratio]{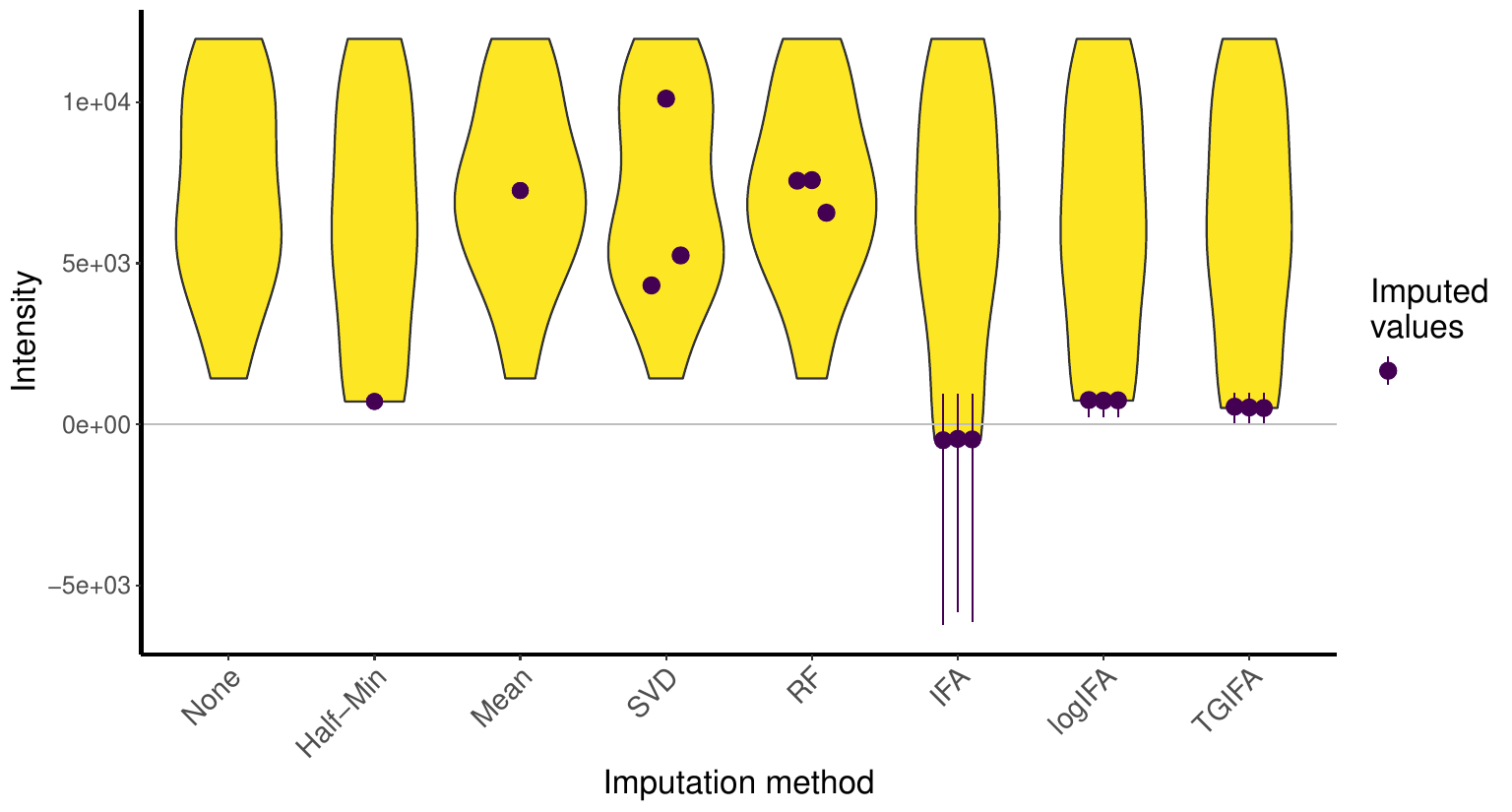}
  \caption{}
\label{fig:broc_91_violin}
\end{subfigure}
\begin{subfigure}{\textwidth}
  \centering
  \includegraphics[width=\textwidth, keepaspectratio]{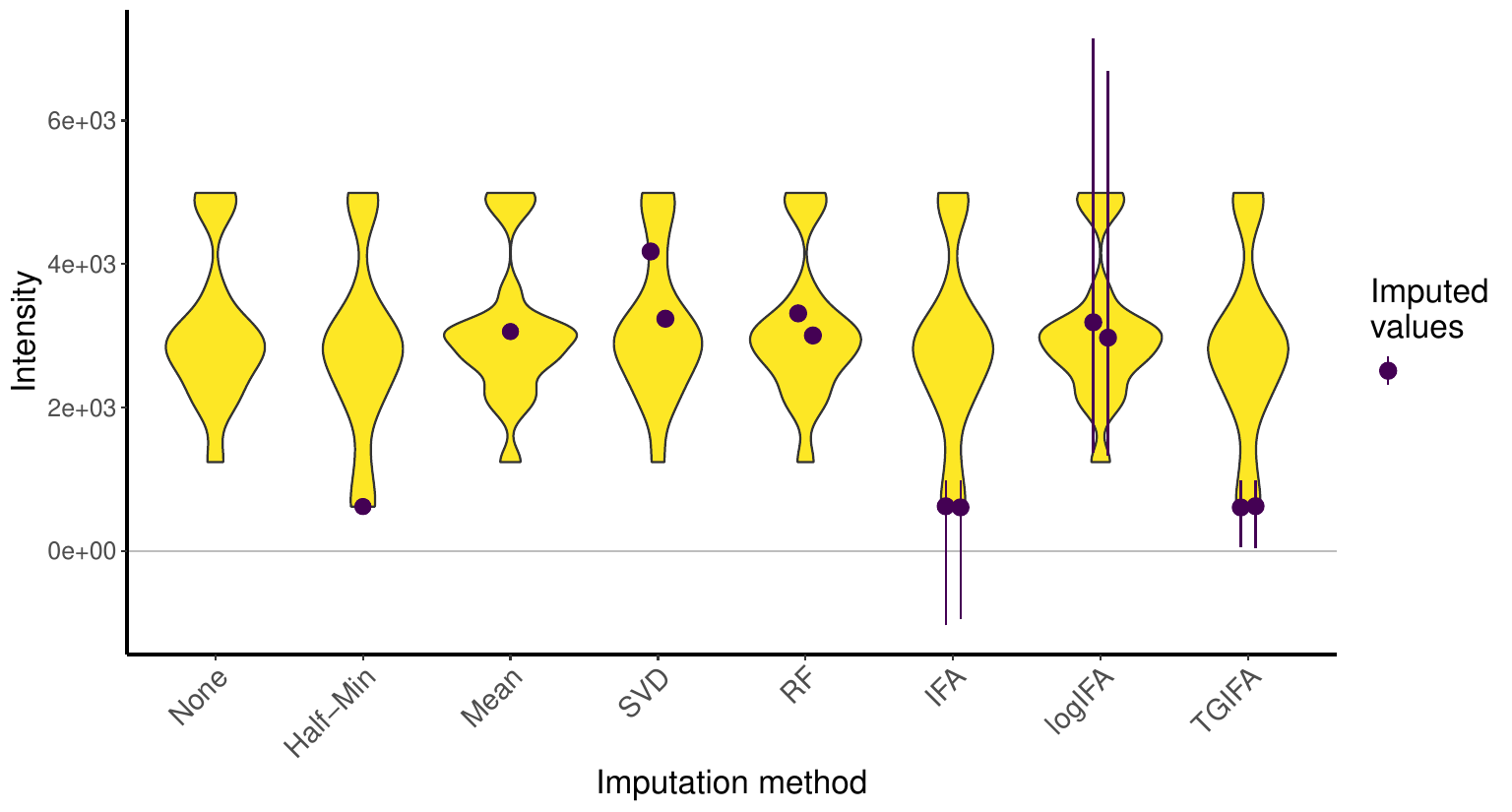}
  \caption{}
\label{fig:broc_4_violin}
\end{subfigure}
\caption{Violin plots and imputed values for two variables with low means and variances from the urinary metabolomics dataset: (a) variable 91 with three missing values and (b) variable 4 with two missing values. For half-minimum and mean imputation, only one imputed point is visible, as all missing entries are imputed with the same value. Imputation methods which impute different values for missing entries in the same variable are jittered for clarity, and $95\%$ credible intervals are shown where available.}
\label{fig:violin_91_4}
\end{figure}

A second example, where variables have relatively low means and variances, such that MNAR missingness is likely to be present, is presented in Figure \ref{fig:violin_91_4} for variable 91 (Figure \ref{fig:broc_91_violin}) and variable 4 (Figure \ref{fig:broc_4_violin}), with three and two missing values respectively. The fixed-value imputation methods again have limited utility, while the IFA approach imputes non-meaningful negative values. The RF method imputes plausible values, however, is restricted to the observed domain of the data and therefore does not allow for MNAR missingness. While the posterior median imputed values under the logIFA method are reasonable, for variable 4 logIFA again imputes over-inflated values resulting in a credible interval upper bound much larger than the largest observed value. The TGIFA model imputes all missing entries in both these variables as MNAR, with posterior probabilities of 0.620, 0.631, and 0.627 for variable 91 and 0.746 and 0.775 for variable 4, with meaningful posterior median imputed values, and credible intervals.

\begin{figure}[htp]
\begin{subfigure}{\textwidth}
  \centering
  \includegraphics[width=\textwidth, keepaspectratio]{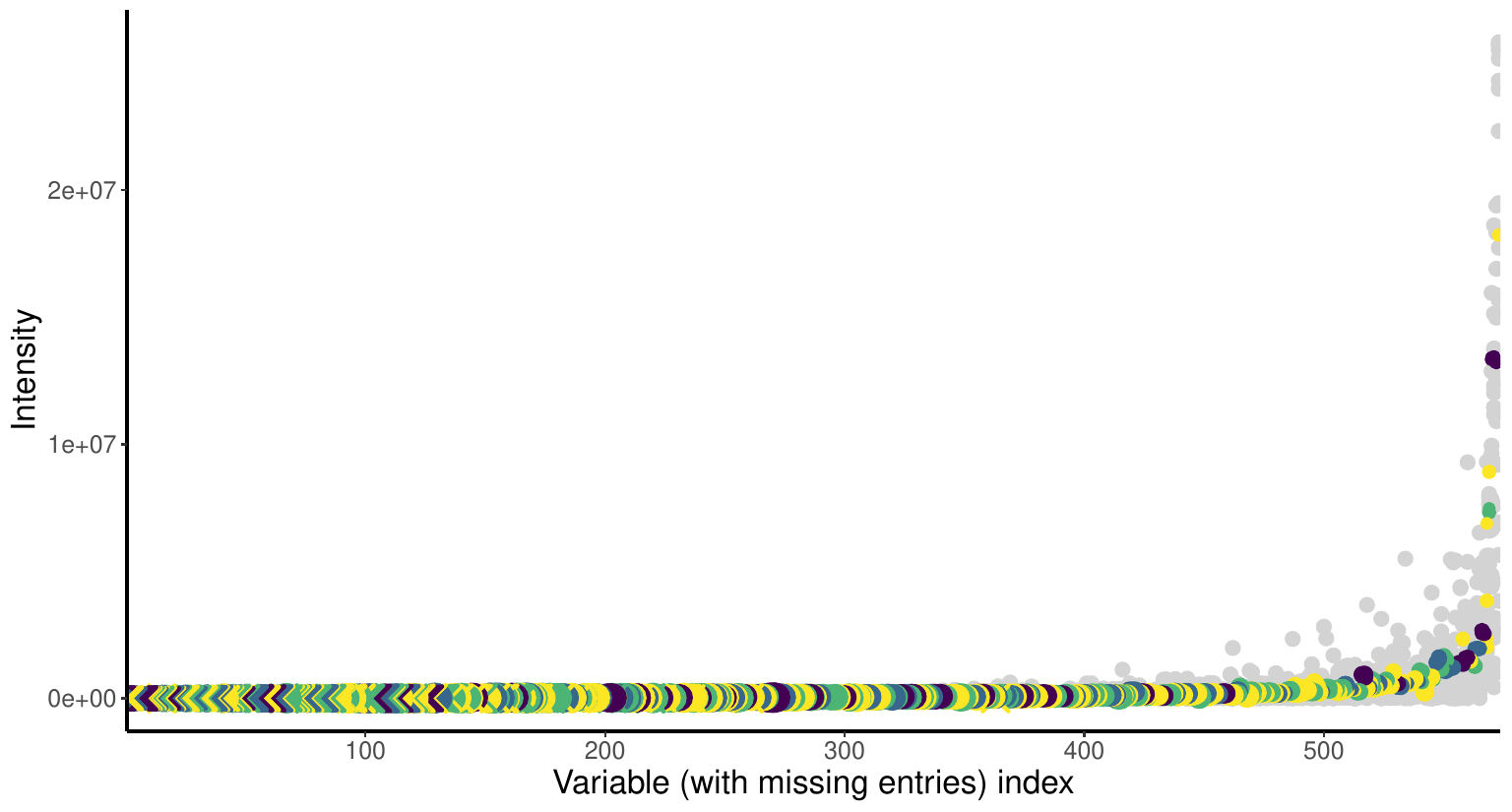}
  \caption{}
  \label{fig:full_imp}
\end{subfigure}
\begin{subfigure}{\textwidth}
  \centering
  \includegraphics[width=\textwidth, keepaspectratio]{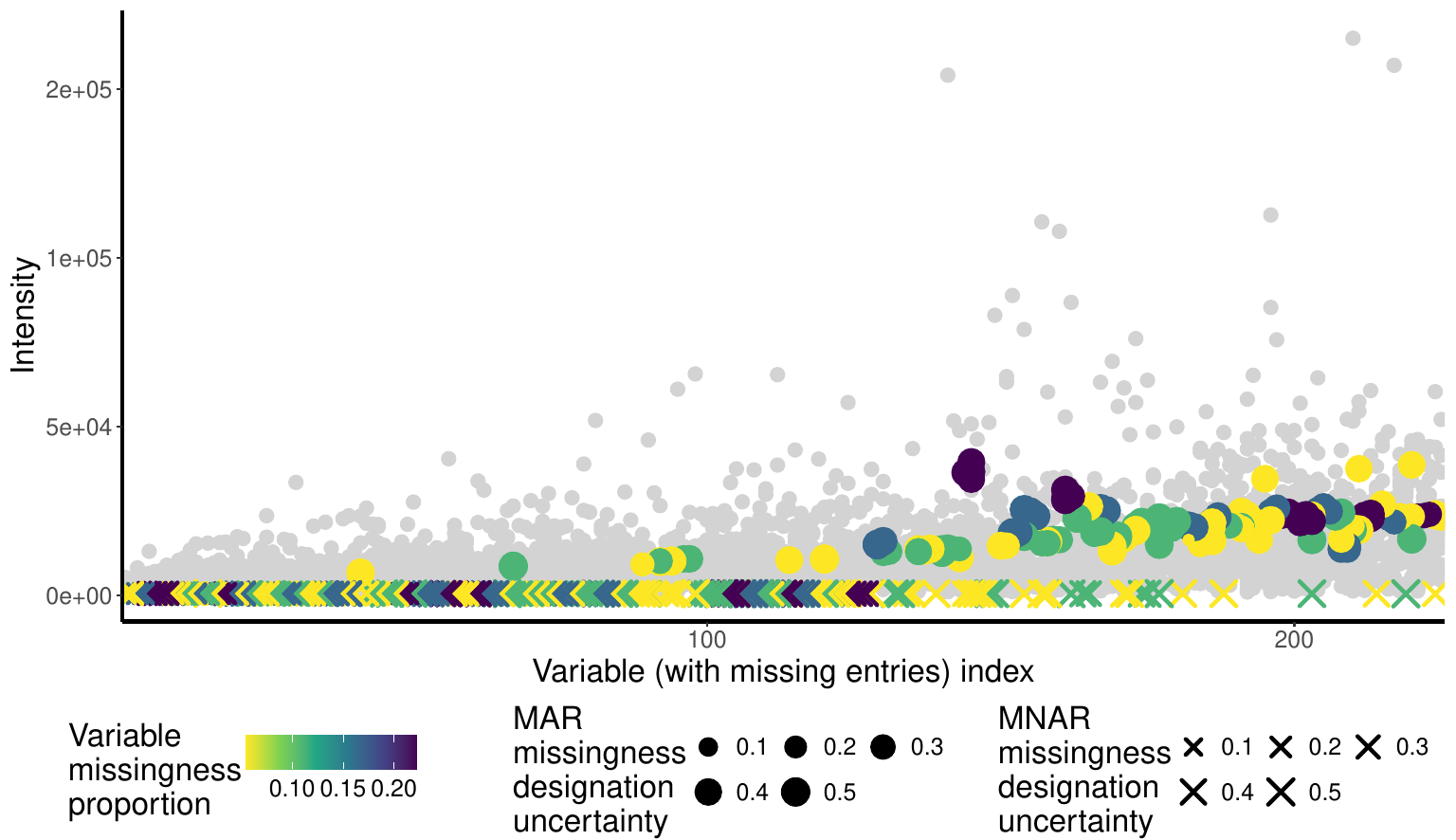}
  \caption{}
  \label{fig:zoomed_imp}
\end{subfigure}
  \caption{Observed and TGIFA-imputed values for the $573$ variables with missing values in the urinary metabolomics dataset. Observed values are shown in grey.
  All 573 variables with missing entries are shown in (a), while (b) shows a zoomed view of 225 earlier variables (with missingness) given their lower observed values and proximity to the LOD.}
  \label{fig:full_data_unc_fig}
\end{figure}

Finally, Figure \ref{fig:full_data_unc_fig} shows imputed and observed values for the variables with missingness. All variables with missing entries are presented in Figure \ref{fig:full_imp} while Figure \ref{fig:zoomed_imp} only shows the 225 variables with the lowest observed means to provide a clearer view of their mixed imputation types and varying designation uncertainties. 
In general, for variables with larger observed means and variances the TGIFA method designates missing values as MAR and imputes plausible values within the range of the observed values; this is intuitive as the observed values are not close to the LOD. In variables with lower observed means and variances (Figure \ref{fig:zoomed_imp}), the TGIFA model infers a mix of both MAR and MNAR-designated imputations. On average, the TGIFA model imputes MAR-designated entries with lower designation-uncertainty than MNAR values ($0.208$ versus $0.356$, respectively). However, the variability in designation-uncertainty is higher in MAR-designated entries compared to MNAR-designated entries (standard deviations of $0.150$ versus $0.086$, respectively). This is intuitive, as MNAR values are not likely to be present in variables with observed values far from the LOD, whereas MAR values are likely to appear across all variables in the dataset.
In summary, while the true values of missing entries are unknown here, imputation using the TGIFA approach allows for both types of missingness, results in physically plausible imputed values, and provides uncertainty quantification.

\section{Discussion}
\label{section:discussion}

Addressing the issue of missing data in high-dimensional metabolomics data is important as data acquisition can be difficult and expensive and many commonly used downstream analysis methods require a complete dataset. The proposed TGIFA approach provides a statistically principled approach to imputing such missing data: it ensures that different types of missingness are accounted for, that the resulting imputed values have meaning for the user, and that the inherent uncertainty in both the imputed values and the missing data type is available. Further, TGIFA provides a parsimonious model for metabolomics data that accounts for its typical $n \ll p$ dimensionality and high levels of dependence between variables. Comprehensive simulation studies demonstrate the performance of TGIFA; physically meaningful imputed values are provided, and the advantages of TGIFA are emphasised in comparison to existing imputation methods. Application of TGIFA to a urinary LC-MS metabolomics dataset highlights its utility, as imputation via TGIFA results in a full, practically useful dataset, with associated uncertainty provided, available for subsequent metabolomics analysis. The provision of open-source R code facilitates use of TGIFA more broadly. 

Truncated distributions, particularly in a multivariate setting, can be cumbersome and time-consuming to work with due to the need to evaluate high-dimensional normalisation integrals. While theoretically tractable, their evaluation can threaten the computational viability of inferential procedures, particularly in MCMC settings. The novel use of the exchange algorithm here for inference on the TGIFA model obviates the need to evaluate these integrals, facilitating the practical use of truncated distributions in an MCMC setting. The inferential procedure utilised herein, while demonstrated for imputation of missing data, should prove useful in general usage of truncated multivariate distributions.

Though applied in this work to metabolomics data, data acquired through LC-MS are prevalent
and TGIFA could prove a useful imputation method in other fields. Additionally, truncated data are present in different scientific contexts, for example in environmental \citep{Kumari2021} and economics \citep{Istaiteyeh2024} research, and the inferential procedure underpinning TGIFA could be utilised for analysis, even if imputation is not the goal of such research.

The TGIFA approach could be extended in several ways. The proposed model assumes data are jointly distributed with the factor scores according to a truncated Gaussian distribution, but different distributional assumptions could be used in order to increase flexibility and account for the typically heavier tailed metabolomics data. The use of truncated versions of the multivariate $t$ \citep{Lee2022} or multivariate normal inverse Gaussian \citep{Barndorff-Nielsen1997, OHaganEtAl2016} distributions, for example, within the TGIFA framework are currently under investigation. 

While a multiplicative truncated gamma process shrinkage prior was employed here to obviate the need to fit multiple models and use selection criteria to choose the optimal model, there are several alternative shrinkage priors that may also be useful. For example, Indian buffet process priors \citep{Knowles2011} and spike-and-slab priors \citep{Legramanti2020} are possible alternatives. Additional research into the properties of factor analysis models in the truncated setting would also be welcome. For example, given the focus of this work on useful imputation model identifiability was not a concern, however further research using approaches that have been considered for standard Gaussian factor analysis (e.g., \cite{Fruhwirth-Schnatter2024}) could be of interest in the truncated setting.
In general, but particularly for these suggested potential extensions, TGIFA would benefit from increased computational efficiency. Alternative, more computationally efficient inferential approaches could be investigated, for example, through the use of variational \citep{Jordan1999} or Hamiltonian Monte Carlo \citep{Duane1987} methods. Finally, TGIFA utilises credible intervals as a means of quantifying the uncertainty of an imputed missing value. Other quantification methods are possible, such as those used for multiple imputation \citep{VanBuuren2021}.

\section{Acknowledgements}
The authors would like to thank Dr Szymon Urbas for useful discussions which contributed to this work.\\

\noindent This publication has emanated from research conducted with the financial support of Taighde Éireann – Research Ireland under grant number 18/CRT/6049 and the  Taighde Éireann – Research Ireland Insight Research Centre under grant number SFI/12/RC/2289\_P2. For the purpose of Open Access, the author has applied a CC BY public copyright licence to any Author Accepted Manuscript version arising from this submission 

\bibliographystyle{apalike}
\bibliography{TGIFA_arXiv}

\appendix

\section{Inferential procedure for TGIFA}
\label{app:FIP}

The full inferential procedure for each parameter of the TGIFA model is detailed below.  
Note that in all cases, $\operatorname{Ga}(\alpha, \beta)$ refers to the gamma distribution whose mean is given by ${\alpha}/{\beta}$. The notation $\operatorname{p}(\theta \mid \dots)$ refers to the conditional posterior distribution of $\theta$ given all other model parameters, and $\operatorname{q}(\breve{\theta} \mid \theta)$ refers to the proposal distribution to propose a new value $\breve{\theta}$ given the current value of the parameter $\theta$. Here, $k^{*}$ is used as a finite, upper bound to the number of latent factors. Note that, given (\ref{equ:y_r_joint}), for all parameters other than $\alpha$, the update steps proceed independently of the value of $r_{ij}$. 

Derivations for the expressions herein are provided, with the exception of the full conditional distributions for $\boldsymbol{\phi}$ and $\boldsymbol{\delta}$ which are unchanged from \cite{Bhattacharya2011, Durante2017}. Note that for clarity, where relevant to parameters that are updated using the exchange algorithm, we refer to the computationally expensive integrals as `intractable'. This slight abuse of nomenclature is to remain in line with literature on the exchange algorithm.

\subsection{Metropolis-Hastings acceptance probability for the latent factor scores}

In the case of the latent factor scores, we find that the standard MH algorithm proves more useful than the exchange algorithm due to the nature of the marginal distribution of $\boldsymbol{\eta}_i^t$. The exchange algorithm, though providing a tractable acceptance probability, proves computationally inefficient as evaluation of the density of a truncated multivariate normal distribution is necessary. The standard MH algorithm, however, with some algebra, does not necessitate this calculation.

The MH acceptance probability, $\mathcal{A}_{\boldsymbol{\eta}_i^{t}}$, for the $i^{\text{th}}$ latent factor score is derived as follows. The proposal distribution $\operatorname{q}(\breve{\boldsymbol{\eta}}_i^{t} \mid \boldsymbol{\eta}_i^{t})$ is given as 
\begin{equation*}
    \operatorname{q}(\boldsymbol{\eta}_i^{t} \mid \dots) \sim N_{k^{*}} \big(  \big[ \boldsymbol{\Lambda}^\top \boldsymbol{\Sigma}^{-1} \boldsymbol{\Lambda} + \boldsymbol{\operatorname{I}}_{k^{*}} \big]^{-1} \boldsymbol{\Lambda}^\top \boldsymbol{\Sigma}^{-1} (\boldsymbol{y}_i^{t} - \boldsymbol{\mu}), \big[ \boldsymbol{\Lambda}^\top \boldsymbol{\Sigma}^{-1} \boldsymbol{\Lambda} + \boldsymbol{\operatorname{I}}_{k^{*}} \big]^{-1}\big),
\end{equation*}
which corresponds to the full conditional distribution for $\boldsymbol{\eta}_i$ in the standard IFA model.

The acceptance probability is then given by
\begin{equation*}
\mathcal{A}_{\boldsymbol{\eta}_i^{t}} = 
    \dfrac{q\left(\boldsymbol{\eta}_i^{t} \mid \breve{\boldsymbol{\eta}}_i^{t}\right) p\left(\breve{\boldsymbol{\eta}}_i^{t}\right) p\left(\boldsymbol{y}_i^{t} \mid \breve{\boldsymbol{\eta}}_i^{t}\right)}{q\left(\breve{\boldsymbol{\eta}}_i^{t} \mid \boldsymbol{\eta}_i^{t}\right) p(\boldsymbol{\eta}_i^{t}) p(\boldsymbol{y}_i^{t} \mid \boldsymbol{\eta}_i^{t})}. 
\end{equation*}
The $\operatorname{p}(\boldsymbol{\eta_i^{t}})$ terms are problematic to evaluate, given their structure as detailed in (\ref{equ:eta_marginal}). Evaluating the denominator of $\operatorname{p}(\boldsymbol{\eta_i^{t}})$ is not required, as the ratio ${\operatorname{p}(\breve{\boldsymbol{\eta}}_i^{t})}/{\operatorname{p}(\boldsymbol{\eta_i^{t}})}$ facilitates its cancellation. However, the numerator of (\ref{equ:eta_marginal}) remains difficult and resource-consuming to evaluate. We therefore  rearrange this integral to facilitate more computationally efficient evaluation as follows, following the procedure to marginalise a joint multivariate Gaussian distribution (see \cite{Bishop2006}). We have that
\begin{equation*}
\begin{array}{rl}
     & \int^{\boldsymbol{\infty}_p}_{\boldsymbol{c}_{\boldsymbol{y}}} \exp \big\{ -\frac{1}{2} [\boldsymbol{y}_i^{t} - \boldsymbol{\mu}, \boldsymbol{\eta}_i^{t}]  \boldsymbol{\Sigma}^{*-1} [\boldsymbol{y}_i^{t} - \boldsymbol{\mu}, \boldsymbol{\eta}_i^{t}]^{\top}\big\} d \boldsymbol{y}_i^{t} \\
     = & \exp \big\{ -\frac{1}{2} \boldsymbol{\eta}_i^{t \top} \boldsymbol{\operatorname{I}}_k \boldsymbol{\eta}_i^{t} \big\} \int^{\boldsymbol{\infty}_p}_{\boldsymbol{c}_{\boldsymbol{y}}} \exp \big\{ -\frac{1}{2} (\boldsymbol{y}_i^{t} - (\boldsymbol{\mu} + \boldsymbol{\Lambda} \boldsymbol{\eta}_i^{t}) )^{\top} \boldsymbol{\Sigma}^{-1}  (\boldsymbol{y}_i^{t} - (\boldsymbol{\mu} + \boldsymbol{\Lambda} \boldsymbol{\eta}_i^{t}) ) \big\} d \boldsymbol{y}_i^{t},
\end{array}
\end{equation*}
and note that the integral is simply the normalisation integral of a truncated multivariate Gaussian distribution $N_p^{[\boldsymbol{c}_{\boldsymbol{y}}, \infty)}(\boldsymbol{y}_i^{t}; \boldsymbol{\mu} + \boldsymbol{\Lambda} \boldsymbol{\eta}_i^{t}, \boldsymbol{\Sigma})$. Thus, the numerator of $\operatorname{p}(\boldsymbol{\eta}_i^{t})$ can be written as
\begin{equation*}
    \dfrac{\exp \big\{ -\frac{1}{2} \big[ \boldsymbol{\eta}_i^{t \top} \boldsymbol{\operatorname{I}}_k \boldsymbol{\eta}_i^{t} + (\boldsymbol{y}_i^{t} - (\boldsymbol{\mu} + \boldsymbol{\Lambda} \boldsymbol{\eta}_i^{t}) )^{\top} \boldsymbol{\Sigma}^{-1}  (\boldsymbol{y}_i^{t} - (\boldsymbol{\mu} + \boldsymbol{\Lambda} \boldsymbol{\eta}_i^{t}) ) \big] \big\} }
    {
     \Bigg[\dfrac{\exp \big\{ -\frac{1}{2} (\boldsymbol{y}_i^{t} - (\boldsymbol{\mu} + \boldsymbol{\Lambda} \boldsymbol{\eta}_i^{t}) )^{\top} \boldsymbol{\Sigma}^{-1}  (\boldsymbol{y}_i^{t} - (\boldsymbol{\mu} + \boldsymbol{\Lambda} \boldsymbol{\eta}_i^{t}) ) \big\} } {
     \int^{\boldsymbol{\infty}_p}_{\boldsymbol{c}_{\boldsymbol{y}}} \exp \big\{ -\frac{1}{2} (\boldsymbol{y}_i^{t} - (\boldsymbol{\mu} + \boldsymbol{\Lambda} \boldsymbol{\eta}_i^{t}) )^{\top} \boldsymbol{\Sigma}^{-1}  (\boldsymbol{y}_i^{t} - (\boldsymbol{\mu} + \boldsymbol{\Lambda} \boldsymbol{\eta}_i^{t}) ) \big\} d \boldsymbol{y}_i^{t}
     } 
     \Bigg]}.
\end{equation*}
This form facilitates convenient cancellation of integrals when considering the form of $p(\boldsymbol{\eta}_i^{t}) p(\boldsymbol{y}_i^{t} \mid \boldsymbol{\eta}_i^{t})$ in $\mathcal{A}_{\boldsymbol{\eta}_i^t}$, resulting in
\begin{equation*}
         p(\boldsymbol{\eta}_i^{t}) p(\boldsymbol{y}_i^{t} \mid \boldsymbol{\eta}_i^{t}) = 
         \dfrac{\exp \big\{ -\frac{1}{2} \big[ \boldsymbol{\eta}_i^{t \top} \boldsymbol{\operatorname{I}}_k \boldsymbol{\eta}_i^{t} + (\boldsymbol{y}_i^{t} - (\boldsymbol{\mu} + \boldsymbol{\Lambda} \boldsymbol{\eta}_i^{t}) )^{\top} \boldsymbol{\Sigma}^{-1}  (\boldsymbol{y}_i^{t} - (\boldsymbol{\mu} + \boldsymbol{\Lambda} \boldsymbol{\eta}_i^{t}) ) \big] \big\} } {\int^{\boldsymbol{\infty}_{p+k}}_{\boldsymbol{c}} \exp \big\{ -\frac{1}{2} [\boldsymbol{y}_i^{t} - \boldsymbol{\mu}, \boldsymbol{\eta}_i^{t}]  \boldsymbol{\Sigma}^{*-1} [\boldsymbol{y}_i^{t} - \boldsymbol{\mu}, \boldsymbol{\eta}_i^{t}]^{\top} \big\} d [\boldsymbol{y}_i^{t}, \boldsymbol{\eta}_i^{t}]^{\top}
         }.
\end{equation*}
This provides a convenient form for the acceptance probability:
\begin{equation*}
    \mathcal{A}_{\boldsymbol{\eta}_i^{t}} = 
    \dfrac{q\left(\boldsymbol{\eta}_i^{t} \mid \breve{\boldsymbol{\eta}}_i^{t}\right) \exp \big\{ -\frac{1}{2} \big[ \breve{\boldsymbol{\eta}}_i^{t \top} \boldsymbol{\operatorname{I}}_k \breve{\boldsymbol{\eta}}_i^{t} + (\boldsymbol{y}_i^{t} - (\boldsymbol{\mu} + \boldsymbol{\Lambda} \breve{\boldsymbol{\eta}}_i^{t}) )^{\top} \boldsymbol{\Sigma}^{-1}  (\boldsymbol{y}_i^{t} - (\boldsymbol{\mu} + \boldsymbol{\Lambda} \breve{\boldsymbol{\eta}}_i^{t}) ) \big] \big\}}{q\left(\breve{\boldsymbol{\eta}}_i^{t} \mid \boldsymbol{\eta}_i^{t}\right) \exp \big\{ -\frac{1}{2} \big[ \boldsymbol{\eta}_i^{t \top} \boldsymbol{\operatorname{I}}_k \boldsymbol{\eta}_i^{t} + (\boldsymbol{y}_i^{t} - (\boldsymbol{\mu} + \boldsymbol{\Lambda} \boldsymbol{\eta}_i^{t}) )^{\top} \boldsymbol{\Sigma}^{-1}  (\boldsymbol{y}_i^{t} - (\boldsymbol{\mu} + \boldsymbol{\Lambda} \boldsymbol{\eta}_i^{t}) ) \big] \big\}}, 
\end{equation*}
obviating the need to evaluate costly high-dimensional integrals.

\subsection{Exchange algorithm acceptance probability for the mean parameter}

In order to derive the exchange algorithm's acceptance probability for the $\boldsymbol{\mu}$ parameter of the TGIFA model, $\mathcal{A}_{\boldsymbol{\mu}}$, we first define the conditional distribution for $\boldsymbol{\mu}$, including the intractable likelihood, then split the likelihood into the tractable part, $\tilde{p}(\boldsymbol{Y}^{t} \mid \boldsymbol{\mu})$, and the computationally expensive integral denoted $\mathcal{Z}(\boldsymbol{\mu})$, as follows:
\begin{equation*}
\begin{array}{rl}
    \operatorname{p}(\boldsymbol{\mu} \mid \dots) & \propto \operatorname{p} (\boldsymbol{Y}^{t} \mid \boldsymbol{\mu}, \boldsymbol{\eta}, \dots ) \operatorname{p}(\boldsymbol{\mu}) \\
    & \propto \displaystyle \prod^n_{i = 1} \big[ \operatorname{p}(\boldsymbol{y}_i^{t} \mid \boldsymbol{\mu}, \boldsymbol{\eta}_i, \dots ) \big] \operatorname{p}(\boldsymbol{\mu}) \\
    & \propto \displaystyle \prod^n_{i = 1} \big[ \dfrac{\tilde{p}(\boldsymbol{y}_i^{t} \mid \boldsymbol{\mu}, \boldsymbol{\eta}_i, \dots)}{\mathcal{Z}(\boldsymbol{\mu})} \big] \operatorname{p}(\boldsymbol{\mu}) \\
    & \propto \displaystyle \prod^n_{i = 1} \big[
    \dfrac{ \exp \big\{ -\frac{1}{2} (\boldsymbol{y}_i^{t} - (\boldsymbol{\mu} + \boldsymbol{\Lambda} \boldsymbol{\eta}_i^{t})) ^{\top} \boldsymbol{\Sigma}^{-1} (\boldsymbol{y}_i^{t} - (\boldsymbol{\mu} + \boldsymbol{\Lambda} \boldsymbol{\eta}_i^{t}))\big\}}{\int^{\boldsymbol{\infty}_p}_{\boldsymbol{c}_{\boldsymbol{y}}} \exp \big\{ -\frac{1}{2} (\boldsymbol{y}_i^{t} - (\boldsymbol{\mu} + \boldsymbol{\Lambda} \boldsymbol{\eta}_i^{t})) ^{\top} \boldsymbol{\Sigma}^{-1} (\boldsymbol{y}_i^{t} - (\boldsymbol{\mu} + \boldsymbol{\Lambda} \boldsymbol{\eta}_i^{t}))\big\} d \boldsymbol{y}_i^{t}} \big] \operatorname{p}(\boldsymbol{\mu}),
\end{array}
\end{equation*}
where $\operatorname{p}(\boldsymbol{\mu}) = N_p(\boldsymbol{\tilde{\mu}}, \boldsymbol{\varphi}^{-1} \boldsymbol{\operatorname{I}}_p)$. We then define a proposal distribution
\begin{equation*}
    \operatorname{q}(\boldsymbol{\mu} \mid \dots ) \sim N_p \Big( \big[n \boldsymbol{\Sigma}^{-1} + \boldsymbol{\varphi} \boldsymbol{\operatorname{I}}_p) \big]^{-1} \big[ \boldsymbol{\Sigma}^{-1} \sum^n_{i=1} (\boldsymbol{y}_i^{t} - \boldsymbol{\Lambda} \boldsymbol{\eta}_i^{t} ) + \boldsymbol{\varphi} \boldsymbol{\operatorname{I}}_p \boldsymbol{\tilde{\mu}} \big], \big[n \boldsymbol{\Sigma}^{-1} + \boldsymbol{\varphi} \boldsymbol{\operatorname{I}}_p) \big]^{-1} \Big),
\end{equation*}
which corresponds to the full conditional distribution for $\boldsymbol{\mu}$ in the standard IFA model.

Thus, following the exchange algorithm and drawing an auxiliary observation $\breve{\boldsymbol{Y}}^t \sim p(\boldsymbol{Y}^{t} \mid \breve{\boldsymbol{\mu}}, \boldsymbol{\eta})$, the following acceptance probability can be derived for a proposed value $\breve{\boldsymbol{\mu}}$ given the current value of $\boldsymbol{\mu}$:
\begin{equation*}
\begin{array}{rl}
    \mathcal{A}_{\boldsymbol{\mu}} = &
    \dfrac{q\left(\boldsymbol{\mu} \mid \breve{\boldsymbol{\mu}}\right) p\left(\breve{\boldsymbol{\mu}}\right) p\left(\boldsymbol{Y}^{t} \mid \breve{\boldsymbol{\mu}}, \boldsymbol{\eta}\right)}{q\left(\breve{\boldsymbol{\mu}} \mid \boldsymbol{\mu}\right) p(\boldsymbol{\mu}) p(\boldsymbol{Y}^{t} \mid \boldsymbol{\mu}, \boldsymbol{\eta})} \dfrac{p(\breve{\boldsymbol{Y}}^t \mid \boldsymbol{\mu}, \boldsymbol{\eta})}{p(\breve{\boldsymbol{Y}}^t \mid \breve{\boldsymbol{\mu}}, \boldsymbol{\eta})} \\ 
    = &
    \dfrac{q\left(\boldsymbol{\mu} \mid \breve{\boldsymbol{\mu}}\right) p\left(\breve{\boldsymbol{\mu}}\right) \tilde{p}\left(\boldsymbol{Y}^{t} \mid \breve{\boldsymbol{\mu}}, \boldsymbol{\eta}\right)}{q\left(\breve{\boldsymbol{\mu}} \mid \boldsymbol{\mu}\right) p(\boldsymbol{\mu}) \tilde{p}(\boldsymbol{Y}^{t} \mid \boldsymbol{\mu}, \boldsymbol{\eta})} \dfrac{\tilde{p}(\breve{\boldsymbol{Y}}^t \mid \boldsymbol{\mu}, \boldsymbol{\eta})}{\tilde{p}(\breve{\boldsymbol{Y}}^t \mid \breve{\boldsymbol{\mu}}, \boldsymbol{\eta})}. \\
    \end{array}
\end{equation*}

\subsection{Exchange algorithm acceptance probability for a row of the loadings matrix}

The exchange algorithm acceptance probability, $\mathcal{A}_{\boldsymbol{\lambda}_j}$, for a row of the loadings matrix, $\boldsymbol{\lambda}_j$, is derived as follows. The proposal distribution $\operatorname{q}(\breve{\boldsymbol{\lambda}}_j \mid \boldsymbol{\lambda}_j)$ is given as 
\begin{equation*}
    \operatorname{p}(\breve{\boldsymbol{\lambda}}_j \mid \dots) \sim N_{k^{*}} ( \boldsymbol{\operatorname{A}}, \boldsymbol{\operatorname{B}}),
\end{equation*}
where
\begin{equation*}
    \boldsymbol{\operatorname{B}} = \big[ \boldsymbol{\operatorname{D}}_j^{-1} + \sigma_j^{-2} \sum^n_{i=1} \boldsymbol{\eta}_i^{t} \boldsymbol{\eta}_i^{t \top}  \big]^{-1},
\end{equation*}

\begin{equation*}
    \boldsymbol{\operatorname{A}} = \boldsymbol{\operatorname{B}} \big[ \sigma_j^{-2} \sum^n_{i=1} (y_{ij}^{t} - \mu_j ) \boldsymbol{\eta}_i^{t} \big],
\end{equation*}
and with $\boldsymbol{\operatorname{D}}_j^{-1} = \operatorname{diag} (\phi_{j1} \tau_{1}, \dots, \phi_{jk^{*}} \tau_{k^{*}})$. This proposal is chosen to correspond to the full conditional distribution for $\boldsymbol{\lambda}_j$ in the standard IFA model. The full conditional distribution for $\boldsymbol{\lambda}_j$ includes the computationally costly normalisation integral. This likelihood is therefore separated into a tractable part, $\tilde{\operatorname{p}}(\boldsymbol{y}_j^{t} \mid \boldsymbol{\lambda}_j)$, and the computationally costly integrals, $\mathcal{Z}(\boldsymbol{\lambda}_j)$, where $\boldsymbol{y}_j^{t}$ indicates column $j$ of the data. The full conditional of $\boldsymbol{\lambda}_j$ is therefore
\begin{equation*}
\begin{array}{rl}
    \operatorname{p}(\boldsymbol{\lambda}_j \mid \dots) & \propto \operatorname{p} (\boldsymbol{y}_j^{t} \mid \boldsymbol{\lambda}_j, \boldsymbol{\eta}, \dots ) \operatorname{p}(\boldsymbol{\lambda}_j \mid \boldsymbol{\phi}_j, \boldsymbol{\tau}) \\
    & \propto \displaystyle \prod^n_{i = 1} \big[ \operatorname{p}(y_{ij}^{t} \mid \boldsymbol{\lambda}_j, \boldsymbol{\eta}_i, \dots ) \big] \operatorname{p}(\boldsymbol{\lambda}_j \mid \boldsymbol{\phi}_j, \boldsymbol{\tau}) \\
    & \propto \displaystyle \prod^n_{i = 1} \big[ \dfrac{\tilde{p}(y_{ij}^{t} \mid \boldsymbol{\lambda}_j, \boldsymbol{\eta}_i)}{\mathcal{Z}(\boldsymbol{\lambda}_j)} \big] \operatorname{p}(\boldsymbol{\lambda}_j \mid \boldsymbol{\phi}_j, \boldsymbol{\tau}) \\
    & \propto \displaystyle \prod^n_{i = 1} \big[
   \dfrac{\psi(\sigma_j^{-1}(y_{ij}^t - (\mu_j + \boldsymbol{\lambda}_j^{\top} \boldsymbol{\eta}_i^{t}))}{\sigma_j (1 - \Phi(\sigma_j^{-1} (0 - (\mu_j + \boldsymbol{\lambda}_j^{\top} \boldsymbol{\eta}_i^{t})))} \big] N_{k^*}(\boldsymbol{0}_{k^*}, \boldsymbol{D}_j), \\
\end{array}
\end{equation*}
with $\psi(\cdot)$ and $\Phi(\cdot)$ representing the probability distribution function and cumulative distribution function respectively of a standard univariate Gaussian distribution. Thus,
\begin{equation*}
\begin{array}{rl}
    \mathcal{A}_{\boldsymbol{\lambda}_j} & = 
    \dfrac{q(\boldsymbol{\lambda}_j \mid \breve{\boldsymbol{\lambda}}_j) p(\breve{\boldsymbol{\lambda}}_j) p(\boldsymbol{y}_j^{t} \mid \breve{\boldsymbol{\lambda}}_j, \boldsymbol{\eta})}{q(\breve{\boldsymbol{\lambda}}_j \mid \boldsymbol{\lambda}_j) p(\boldsymbol{\lambda}_j) p(\boldsymbol{y}_j^{t} \mid \boldsymbol{\lambda}_j, \boldsymbol{\eta})} \dfrac{p(\breve{\boldsymbol{y}}_j^t \mid \boldsymbol{\lambda}_j, \boldsymbol{\eta})}{p(\breve{\boldsymbol{y}}_j^t \mid \breve{\boldsymbol{\lambda}}_j, \boldsymbol{\eta})} \\
    & =
    \dfrac{q(\boldsymbol{\lambda}_j \mid \breve{\boldsymbol{\lambda}}_j) p(\breve{\boldsymbol{\lambda}}_j) \tilde{p}(\boldsymbol{y}_j^{t} \mid \breve{\boldsymbol{\lambda}}_j, \boldsymbol{\eta})}{q(\breve{\boldsymbol{\lambda}}_j \mid \boldsymbol{\lambda}_j) p(\boldsymbol{\lambda}_j) \tilde{p}(\boldsymbol{y}_j^{t} \mid \boldsymbol{\lambda}_j, \boldsymbol{\eta})} \dfrac{\tilde{p}(\breve{\boldsymbol{y}}_j^t \mid \boldsymbol{\lambda}_j, \boldsymbol{\eta})}{\tilde{p}(\breve{\boldsymbol{y}}_j^t \mid \breve{\boldsymbol{\lambda}}_j, \boldsymbol{\eta})}.
    \end{array}
\end{equation*}

\subsection{Exchange algorithm acceptance probability for the diagonal covariance of the idiosyncratic errors}

The exchange algorithm acceptance probability, $\mathcal{A}_{\sigma_j^{-2}}$, for the $jj^{\text{th}}$ entry of $\boldsymbol{\Sigma}$ is derived as follows. The proposal distribution $\operatorname{q}(\breve{\sigma}_j^{-2} \mid \sigma_j^{-2})$ is given as 
\begin{equation*}
    \operatorname{p}(\sigma_j^{-2} \mid \dots) \sim \operatorname{Ga}\Big(\frac{n}{2} + a_{\sigma}, b_{\sigma} + \frac{1}{2} \sum^n_{i=1} (y_{ij}^{t} - \mu_j - \boldsymbol{\lambda}_j^\top \boldsymbol{\eta}_i^{t} )^2\Big),
\end{equation*}
which corresponds to the full conditional distribution for $\sigma_j^{-2}$ in the standard IFA model. The full conditional posterior distribution for $ \sigma_j^{-2}$ includes the computationally costly likelihood. This likelihood is therefore separated into a tractable part, $\tilde{\operatorname{p}}(\boldsymbol{y}_j^{t} \mid  \sigma_j^{-2},  \boldsymbol{\eta})$, and the computationally costly integrals, $\mathcal{Z}(\sigma_j^{-2})$. The full conditional posterior of $ \sigma_j^{-2}$ is therefore
\begin{equation*}
\begin{array}{rl}
    \operatorname{p}( \sigma_j^{-2} \mid \dots) & \propto \operatorname{p} (\boldsymbol{y}_j^{t} \mid  \sigma_j^{-2},  \boldsymbol{\eta}_i, \dots) \operatorname{p}( \sigma_j^{-2}) \\
    & \propto \displaystyle \prod^n_{i = 1} \big[ \operatorname{p}(y_{ij}^{t} \mid  \sigma_j^{-2},  \boldsymbol{\eta}_i, \dots) \big] \operatorname{p}( \sigma_j^{-2}) \\
    & \propto \displaystyle \prod^n_{i = 1} \big[ \dfrac{\tilde{p}(y_{ij}^{t} \mid  \sigma_j^{-2},  \boldsymbol{\eta}_i, \dots)}{\mathcal{Z}( \sigma_j^{-2})} \big] \operatorname{p}( \sigma_j^{-2}) \\
    & \propto \displaystyle \prod^n_{i = 1} \big[
   \dfrac{\psi(\sigma_j^{-1}(y_{ij}^t - (\mu_j + \boldsymbol{\lambda}_j^{\top} \boldsymbol{\eta}_i^{t}))}{\sigma_j (1 - \Phi(\sigma_j^{-1} (0 - (\mu_j + \boldsymbol{\lambda}_j^{\top} \boldsymbol{\eta}_i^{t})))} \big] \operatorname{Ga}(a_{\sigma}, b_{\sigma}), \\
\end{array}
\end{equation*}
with $\psi(\cdot)$ and $\Phi(\cdot)$ representing the probability distribution function and cumulative distribution function respectively of a univariate Gaussian distribution. Thus, the acceptance probability is given by
\begin{equation*}
\begin{array}{rl}
    \mathcal{A}_{\sigma_j^{-2}} & = 
    \dfrac{q\left(\sigma_j^{-2} \mid \breve{\sigma}_j^{-2}\right) p\left(\breve{\sigma}_j^{-2}\right) p\left(\boldsymbol{y}_j^{t} \mid \breve{\sigma}_j^{-2},  \boldsymbol{\eta}_i \right)}{q\left(\breve{\sigma}_j^{-2} \mid \sigma_j^{-2}\right) p(\sigma_j^{-2}) p(\boldsymbol{y}_j^{t} \mid \sigma_j^{-2},  \boldsymbol{\eta}_i)} \dfrac{p(\breve{\boldsymbol{y}}_j^t \mid \sigma_j^{-2},  \boldsymbol{\eta}_i)}{p(\breve{\boldsymbol{y}}_j^t \mid \breve{\sigma}_j^{-2},  \boldsymbol{\eta}_i)} \\
    & =
    \dfrac{q\left(\sigma_j^{-2} \mid \breve{\sigma}_j^{-2}\right) p\left(\breve{\sigma}_j^{-2}\right) \tilde{p}\left(\boldsymbol{y}_j^{t} \mid \breve{\sigma}_j^{-2},  \boldsymbol{\eta}_i \right)}{q\left(\breve{\sigma}_j^{-2} \mid \sigma_j^{-2}\right) p(\sigma_j^{-2}) \tilde{p}(\boldsymbol{y}_j^{t} \mid \sigma_j^{-2},  \boldsymbol{\eta}_i)} \dfrac{\tilde{p}(\breve{\boldsymbol{y}}_j^t \mid \sigma_j^{-2},  \boldsymbol{\eta}_i)}{\tilde{p}(\breve{\boldsymbol{y}}_j^t \mid \breve{\sigma}_j^{-2},  \boldsymbol{\eta}_i)}.
\end{array}
\end{equation*}

\subsection{The full conditional posterior distribution for the MAR missingness probability}

The $\alpha$ parameter represents the probability that $r_{ij} = 1$ given that $y_{ij} > \text{LOD}$. For each $i = 1, \dots, n$, and $j = 1, \dots, p$:

\begin{equation*}
\begin{array}{rl}
    \operatorname{p}(\alpha \mid \dots ) \propto & \displaystyle \prod_{i=1}^n \prod_{j=1}^p \operatorname{p}(r_{ij} \mid y_{ij}, \boldsymbol{\theta}) \operatorname{p}(\alpha) \\
    \propto & \displaystyle \prod_{i=1}^n \prod_{j=1}^p (1 - \alpha) ^{\mathds{1}\{r_{ij}=1\}} \alpha ^{\mathds{1}\{r_{ij}=0\}} \\
    \propto & (1 - \alpha) ^{\sum_{i=1}^n \sum_{j=1}^p \mathds{1}\{r_{ij}=1\}} \alpha ^{\sum_{i=1}^n \sum_{j=1}^p\mathds{1}\{r_{ij}=0\}} \\
    \propto & (1 - \alpha) ^{N_{o}^{LOD^+}} \alpha ^{N_{m}^{LOD^+}},
\end{array}
\end{equation*}
where $N_{m}^{LOD^+}$ is the number of points inferred to be missing above the LOD, and $N_{o}^{LOD^+}$ is the number of points observed above the LOD. This matches the functional form of a beta distribution, thus
\begin{equation*}
    \operatorname{p}(\alpha \mid \dots ) \propto 
    \operatorname{Beta} (N_{m}^{LOD^+} + 1, N_{o}^{LOD^+} + 1).
\end{equation*}

\subsection{Summary of inferential procedure}

The inferential procedure for TGIFA is summarised as follows. The latent factor scores are updated using a MH step with acceptance probability

\begin{equation*}
    \mathcal{A}_{\boldsymbol{\eta}_i^{t}} = 
    \dfrac{q\left(\boldsymbol{\eta}_i^{t} \mid \breve{\boldsymbol{\eta}}_i^{t}\right) \exp \big\{ -\frac{1}{2} \big[ \breve{\boldsymbol{\eta}}_i^{t \top} \boldsymbol{\operatorname{I}}_k \breve{\boldsymbol{\eta}}_i^{t} + (\boldsymbol{y}_i^{t} - (\boldsymbol{\mu} + \boldsymbol{\Lambda} \breve{\boldsymbol{\eta}}_i^{t}) )^{\top} \boldsymbol{\Sigma}^{-1}  (\boldsymbol{y}_i^{t} - (\boldsymbol{\mu} + \boldsymbol{\Lambda} \breve{\boldsymbol{\eta}}_i^{t}) ) \big] \big\}}{q\left(\breve{\boldsymbol{\eta}}_i^{t} \mid \boldsymbol{\eta}_i^{t}\right) \exp \big\{ -\frac{1}{2} \big[ \boldsymbol{\eta}_i^{t \top} \boldsymbol{\operatorname{I}}_k \boldsymbol{\eta}_i^{t} + (\boldsymbol{y}_i^{t} - (\boldsymbol{\mu} + \boldsymbol{\Lambda} \boldsymbol{\eta}_i^{t}) )^{\top} \boldsymbol{\Sigma}^{-1}  (\boldsymbol{y}_i^{t} - (\boldsymbol{\mu} + \boldsymbol{\Lambda} \boldsymbol{\eta}_i^{t}) ) \big] \big\}}. 
\end{equation*}

The mean parameter is updated using the exchange algorithm with acceptance probability

\begin{equation*}
    \mathcal{A}_{\boldsymbol{\mu}} = 
    \frac{q\left(\boldsymbol{\mu} \mid \breve{\boldsymbol{\mu}}\right) p\left(\breve{\boldsymbol{\mu}}\right) \tilde{p}\left(\boldsymbol{Y}^{t} \mid \breve{\boldsymbol{\mu}}, \boldsymbol{\eta} \right)}{q\left(\breve{\boldsymbol{\mu}} \mid \boldsymbol{\mu}\right) p(\boldsymbol{\mu}) \tilde{p}(\boldsymbol{Y}^{t} \mid \boldsymbol{\mu}, \boldsymbol{\eta})} \frac{\tilde{p}(\breve{\boldsymbol{Y}}^t \mid \boldsymbol{\mu}, \boldsymbol{\eta})}{\tilde{p}(\breve{\boldsymbol{Y}}^t \mid \breve{\boldsymbol{\mu}},\boldsymbol{\eta})}.
\end{equation*}

Each row of the loadings matrix is updated using the exchange algorithm with acceptance probability

\begin{equation*}
    \mathcal{A}_{\boldsymbol{\lambda}_j} = 
    \dfrac{q(\boldsymbol{\lambda}_j \mid \breve{\boldsymbol{\lambda}}_j) p(\breve{\boldsymbol{\lambda}}_j) \tilde{p}(\boldsymbol{y}_j^{t} \mid \breve{\boldsymbol{\lambda}}_j, \boldsymbol{\eta})}{q(\breve{\boldsymbol{\lambda}}_j \mid \boldsymbol{\lambda}_j) p(\boldsymbol{\lambda}_j) \tilde{p}(\boldsymbol{y}_j^{t} \mid \boldsymbol{\lambda}_j, \boldsymbol{\eta})} \dfrac{\tilde{p}(\breve{\boldsymbol{y}}_j^t \mid \boldsymbol{\lambda}_j, \boldsymbol{\eta})}{\tilde{p}(\breve{\boldsymbol{y}}_j^t \mid \breve{\boldsymbol{\lambda}}_j, \boldsymbol{\eta})}.
\end{equation*}

Each element of the diagonal covariance of the idiosyncratic errors is updated using the exchange algorithm with an acceptance probability 

\begin{equation*}
    \mathcal{A}_{\sigma_j^{-2}} = 
    \dfrac{q\left(\sigma_j^{-2} \mid \breve{\sigma}_j^{-2}\right) p\left(\breve{\sigma}_j^{-2}\right) \tilde{p}\left(\boldsymbol{y}_j^{t} \mid \breve{\sigma}_j^{-2}, \boldsymbol{\eta}\right)}{q\left(\breve{\sigma}_j^{-2} \mid \sigma_j^{-2}\right) p(\sigma_j^{-2}) \tilde{p}(\boldsymbol{y}_j^{t} \mid \sigma_j^{-2}, \boldsymbol{\eta})} \dfrac{\tilde{p}(\breve{\boldsymbol{y}}_j^t \mid \sigma_j^{-2}, \boldsymbol{\eta})}{\tilde{p}(\breve{\boldsymbol{y}}_j^t \mid \breve{\sigma}_j^{-2}, \boldsymbol{\eta})}.
\end{equation*}

For the remaining parameters, inference can proceed via a Gibbs sampler as the full conditional posterior distributions are available in closed form as follows:

\begin{equation*}
    \operatorname{p}( \phi_{jh} \mid \dots)\sim \operatorname{Ga} \Bigg(\frac{1}{2} + \kappa_1, \frac{\tau_h \lambda_{jh}^2}{2} + \kappa_2\Bigg),
\end{equation*}

\begin{equation*}
\operatorname{p}(\delta_1 \mid \dots) \sim \operatorname{Ga} \Bigg( a_1 + \frac{pk^{*}}{2}, 1 + \frac{\sum_{l=1}^{k^{*}} \tau_l^{(1)} \sum_{j=1}^{p}  \phi_{jl} \lambda_{jl}^2}{2}\Bigg),
\end{equation*}

\begin{equation*}
\operatorname{p}(\delta_h \mid \dots) \sim \operatorname{Ga}^{[1, \infty)} \Bigg( a_2 + \frac{p ( k^{*} - h + 1)}{2}, 1 + \frac{\sum_{l=h}^{k^{*}} \tau_l^{(h)} \sum_{j=1}^{p} \phi_{jl} \lambda_{jl}^2}{2}\Bigg), \quad h \geq 2,
\end{equation*}
where $\tau_l^{(h)}= \displaystyle \prod_{t=1, t \neq h}^l \delta_t \text { for } h=1, \dots, k^{*}$, and

\begin{equation*}
    \operatorname{p}(\alpha \mid \dots ) \propto 
    \operatorname{Beta} (N_{m}^{LOD^{+}} + 1, N_{o}^{LOD^+} + 1),
\end{equation*}
where $N_{m}^{LOD^+}$ is the number of points inferred to be missing above the LOD and $N_{o}^{LOD+}$ is the number of points observed above the LOD.

\section{Additional simulation study materials}
\label{app:simstudy}

\renewcommand{\thefigure}{C\arabic{figure}}
\setcounter{figure}{0}

\begin{figure}[htp]
\begin{subfigure}{\textwidth}
  \centering \includegraphics[height=0.2\textheight, keepaspectratio]{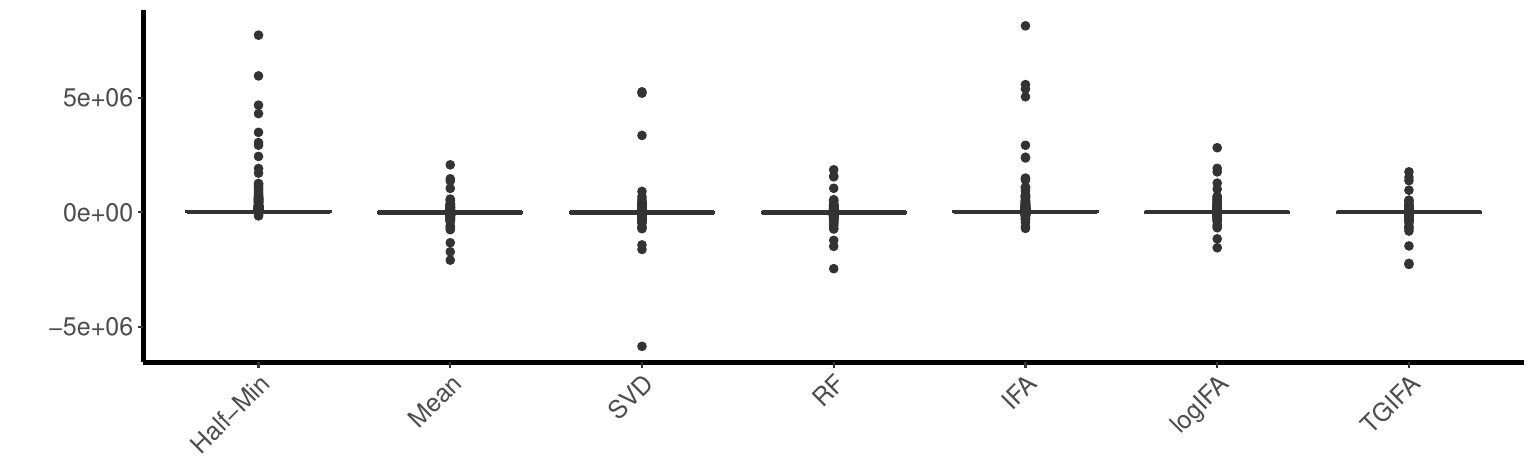}
  \caption{}
   \label{fig:resids_boxplots_all}
\end{subfigure}
\begin{subfigure}{\textwidth}
  \centering
  \includegraphics[height=0.2\textheight, keepaspectratio]{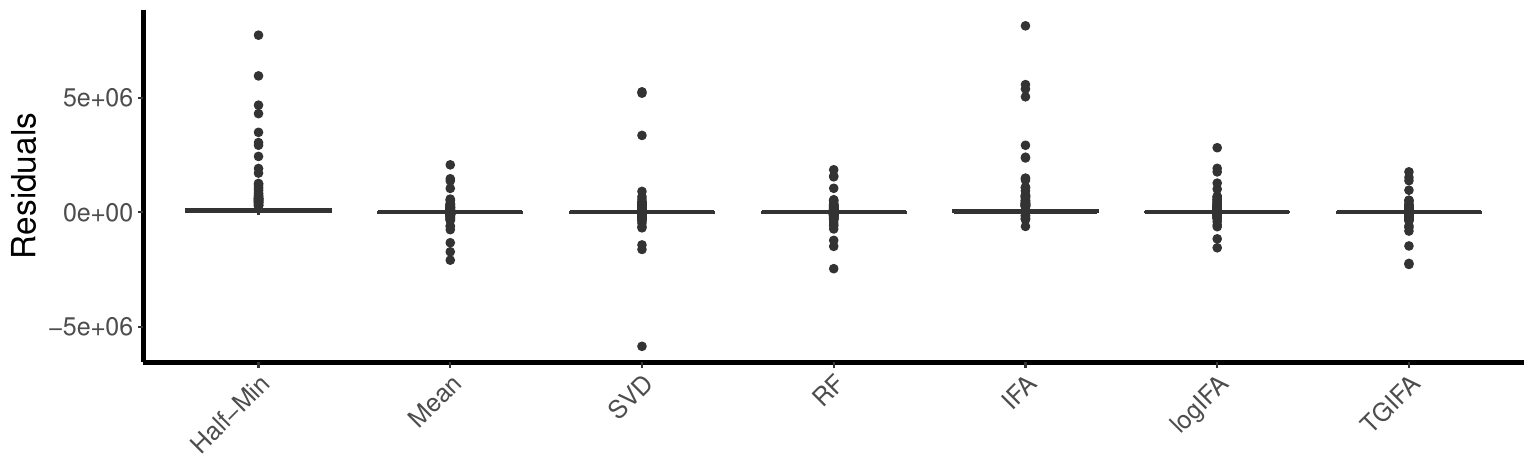}
  \caption{}
   \label{fig:resids_boxplots_MAR}
\end{subfigure}
\begin{subfigure}{\textwidth}
  \centering
  \includegraphics[height=0.2\textheight, keepaspectratio]{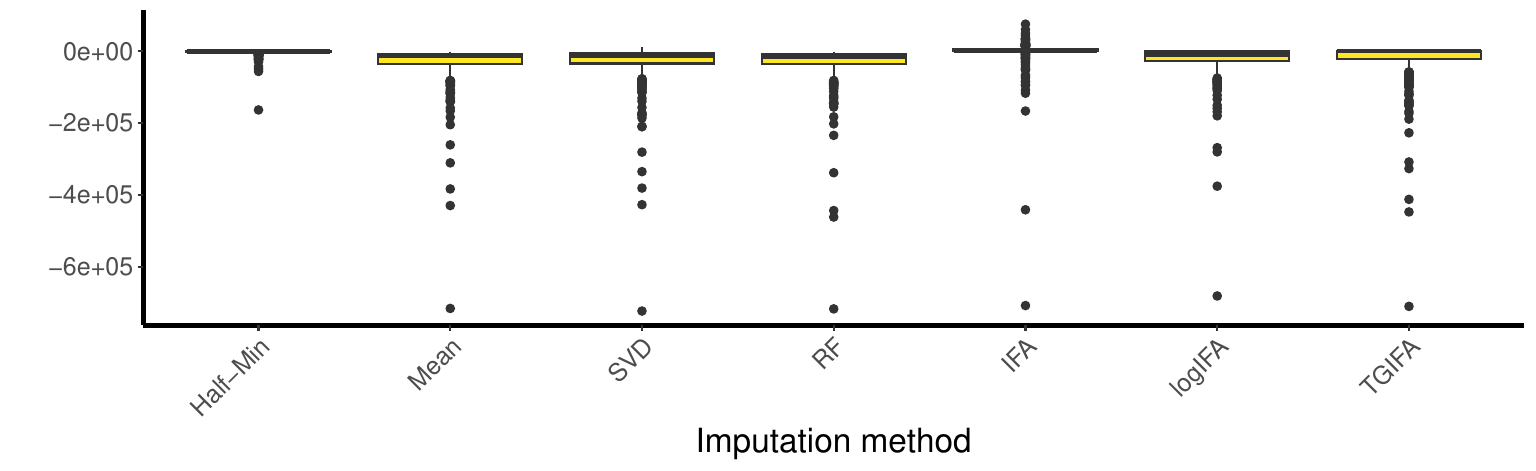}
  \caption{}
   \label{fig:resids_boxplots_MNAR}
\end{subfigure}
  \caption{Residuals between (posterior median) imputed values and true values for (a) all imputed values, (b) MAR imputed values and (c) MNAR imputed values for one simulated dataset. 
  }
  \label{fig:resids_boxplots}
\end{figure}

\begin{figure}[htp]
  \centering
  \includegraphics[width=\textwidth, keepaspectratio]{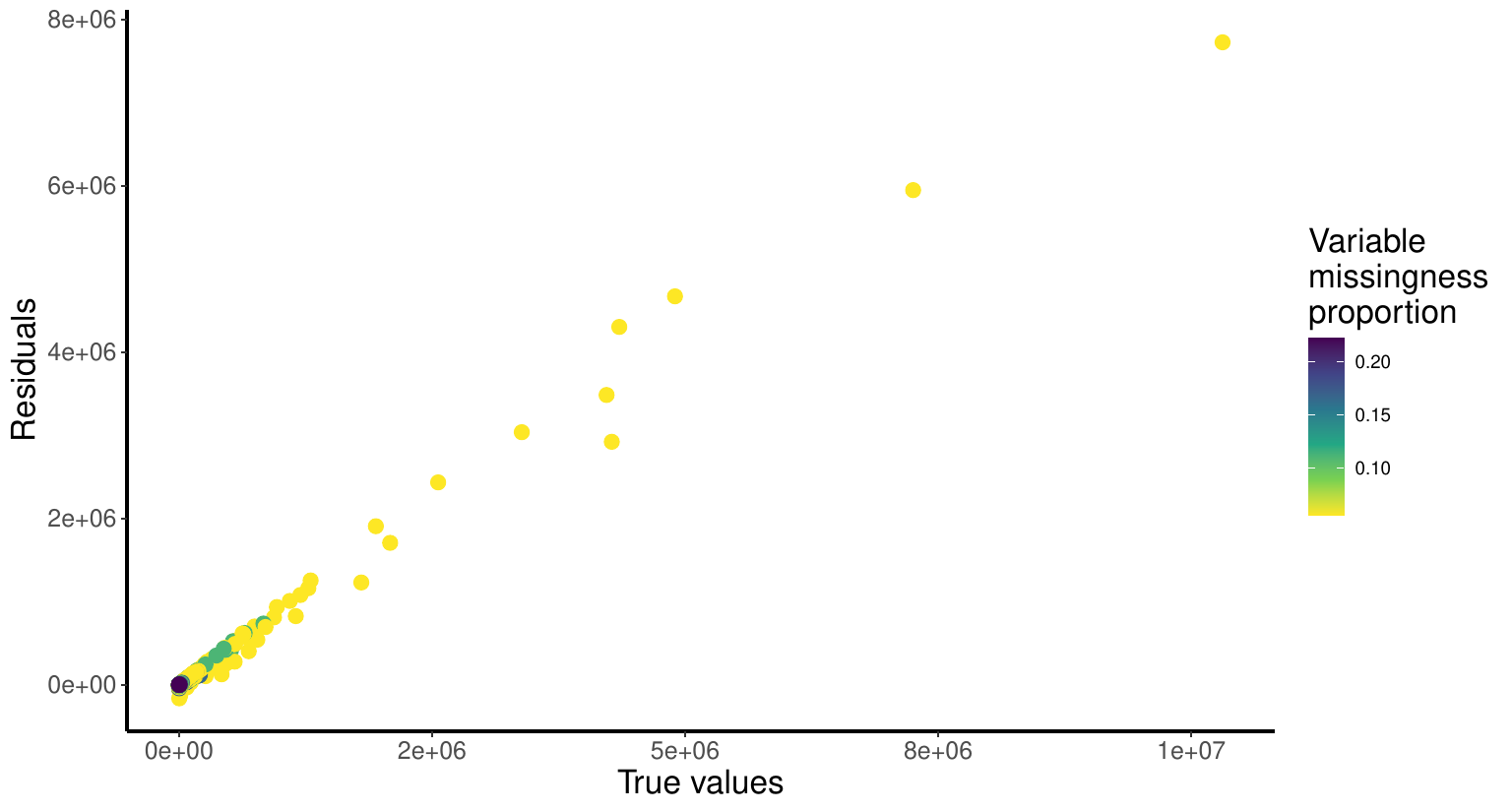}
\caption{Residuals between imputed values and true values under half-minimum imputation for one simulated dataset.}
\label{fig:resids_half_min}
\end{figure}

\begin{figure}[htp]
  \centering
 \includegraphics[width=\textwidth, keepaspectratio]{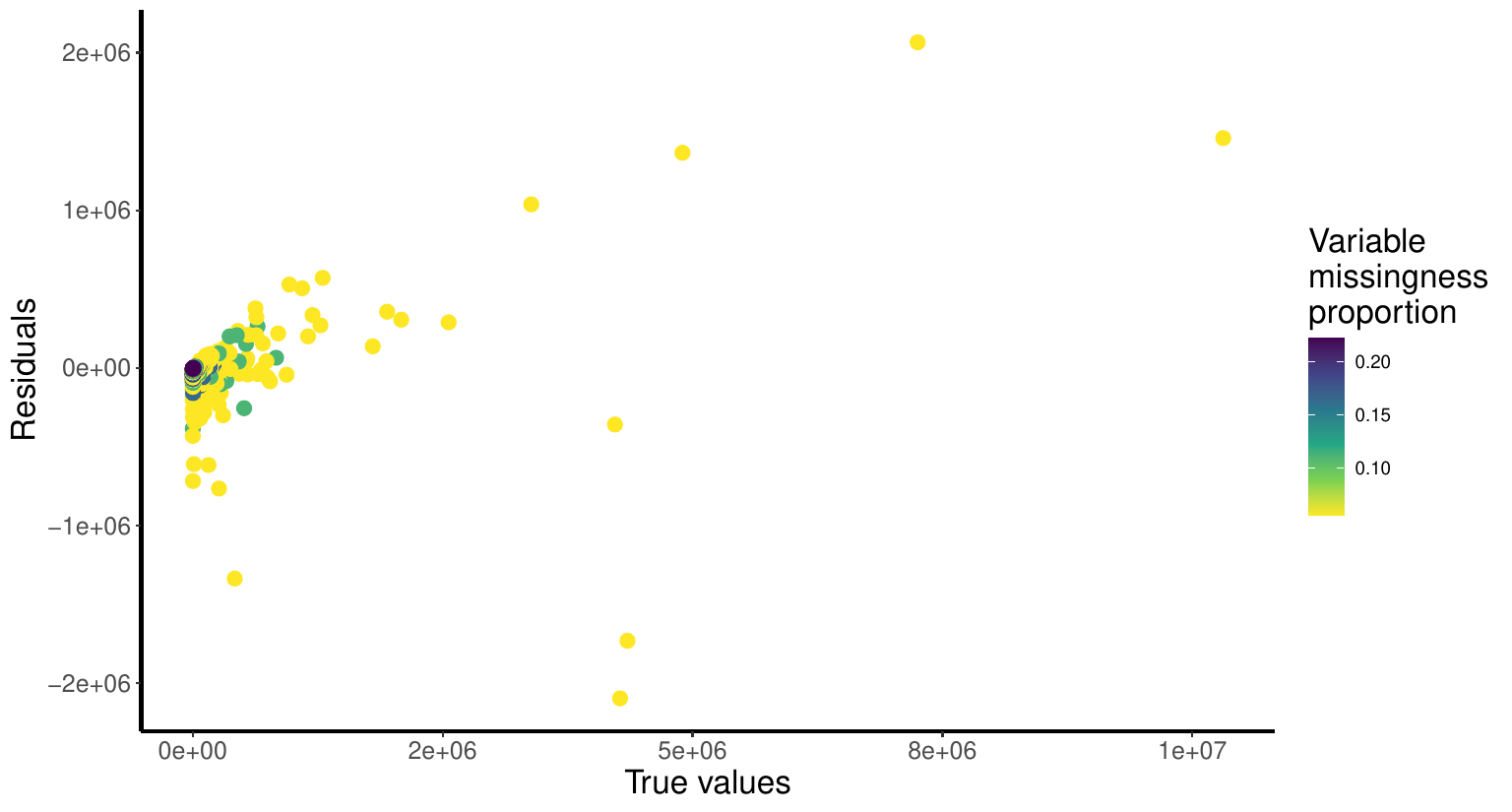}
\caption{Residuals between imputed values and true values under mean imputation for one simulated dataset.}
\label{fig:resids_mean}
\end{figure}

\begin{figure}[htp]
  \centering
  \includegraphics[width=\textwidth, keepaspectratio]{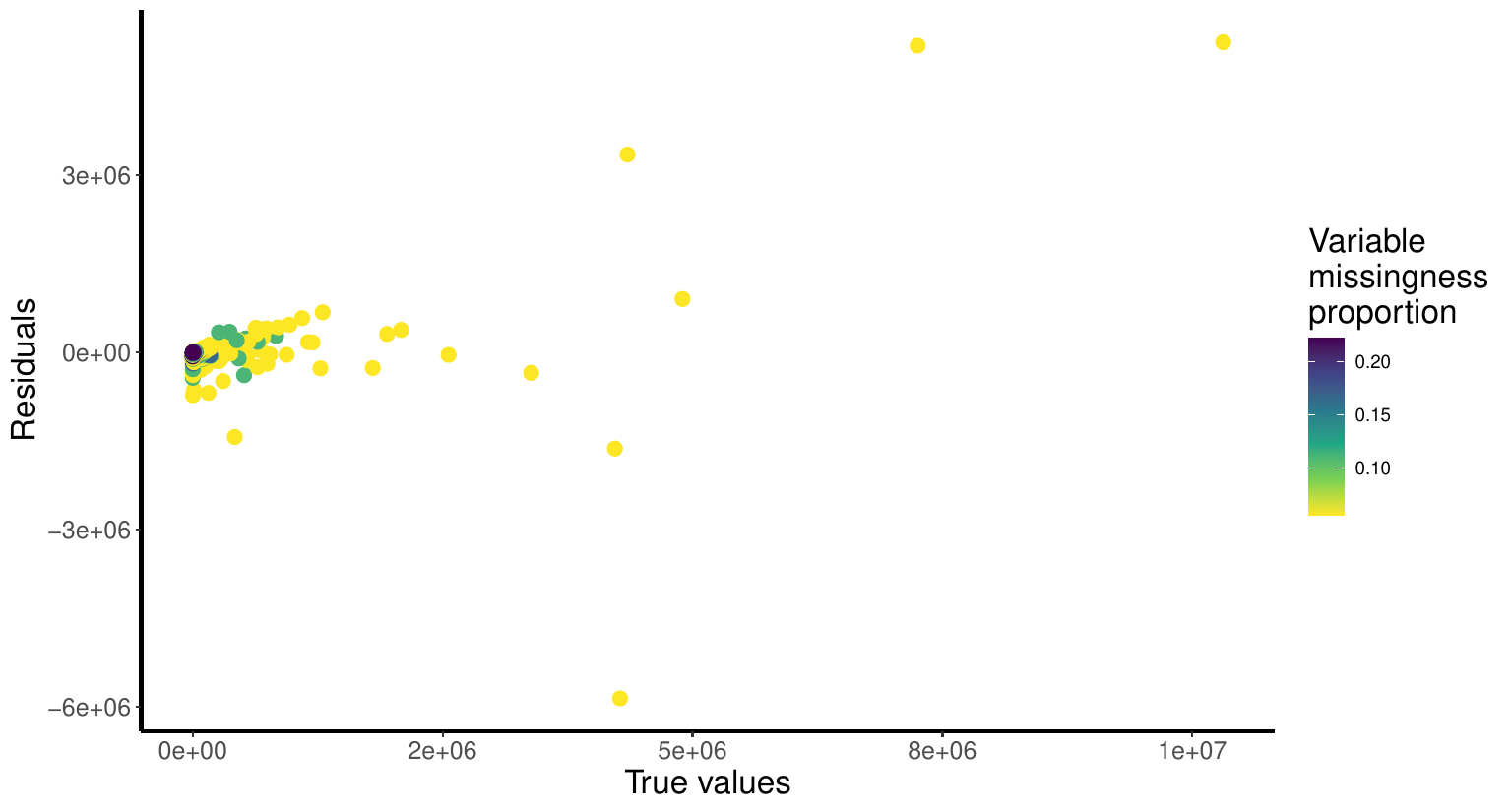}
\caption{Residuals between imputed values and true values under the SVD approach for one simulated dataset.}
\label{fig:resids_svd}
\end{figure}

\begin{figure}[htp]
  \centering
  \includegraphics[width=\textwidth, keepaspectratio]{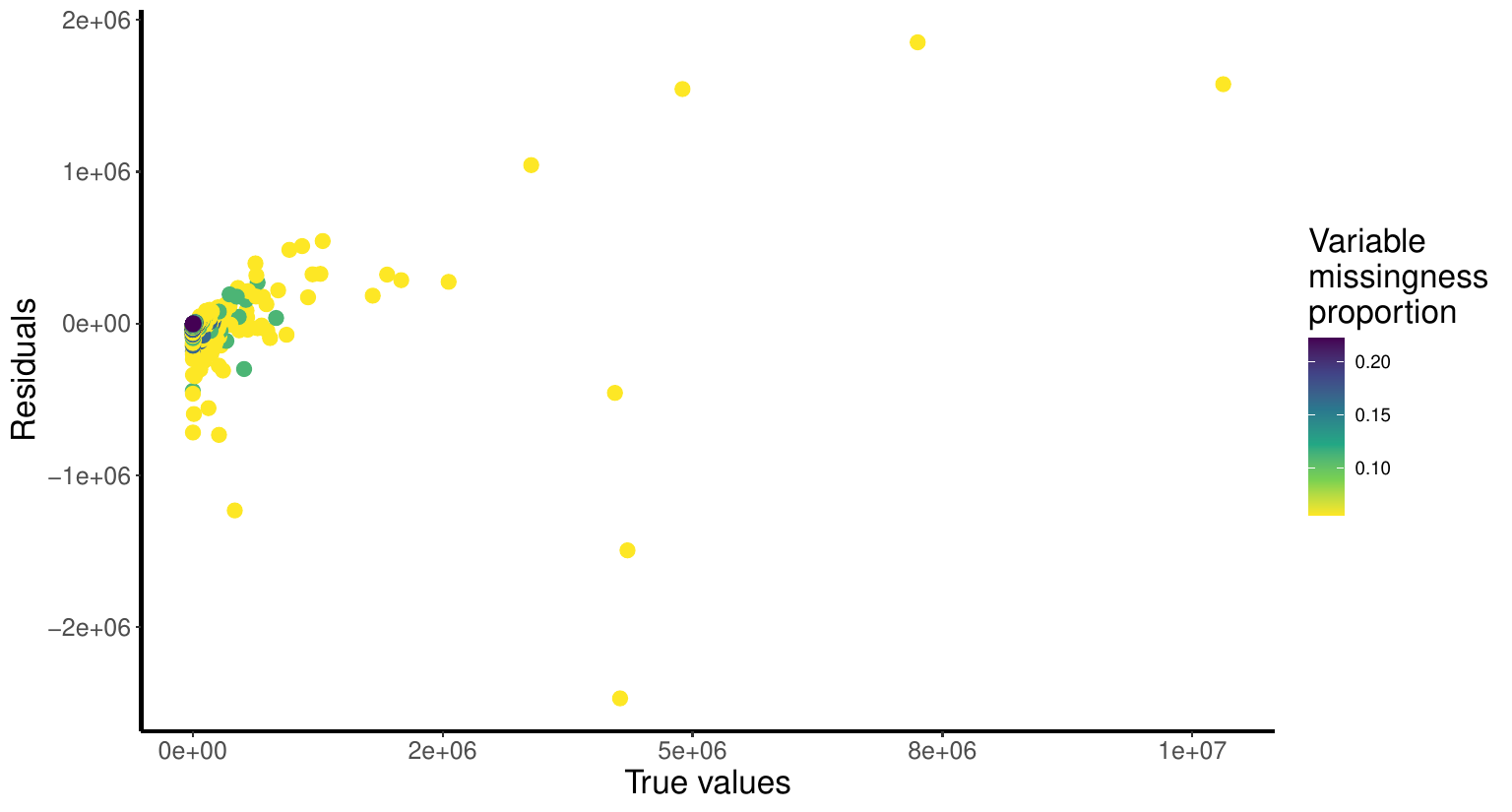}
\caption{Residuals between imputed values and true values under the RF approach for one simulated dataset.}
\label{fig:resids_rf}
\end{figure}

\begin{figure}[htp]
  \centering
  \includegraphics[width=\textwidth, keepaspectratio]{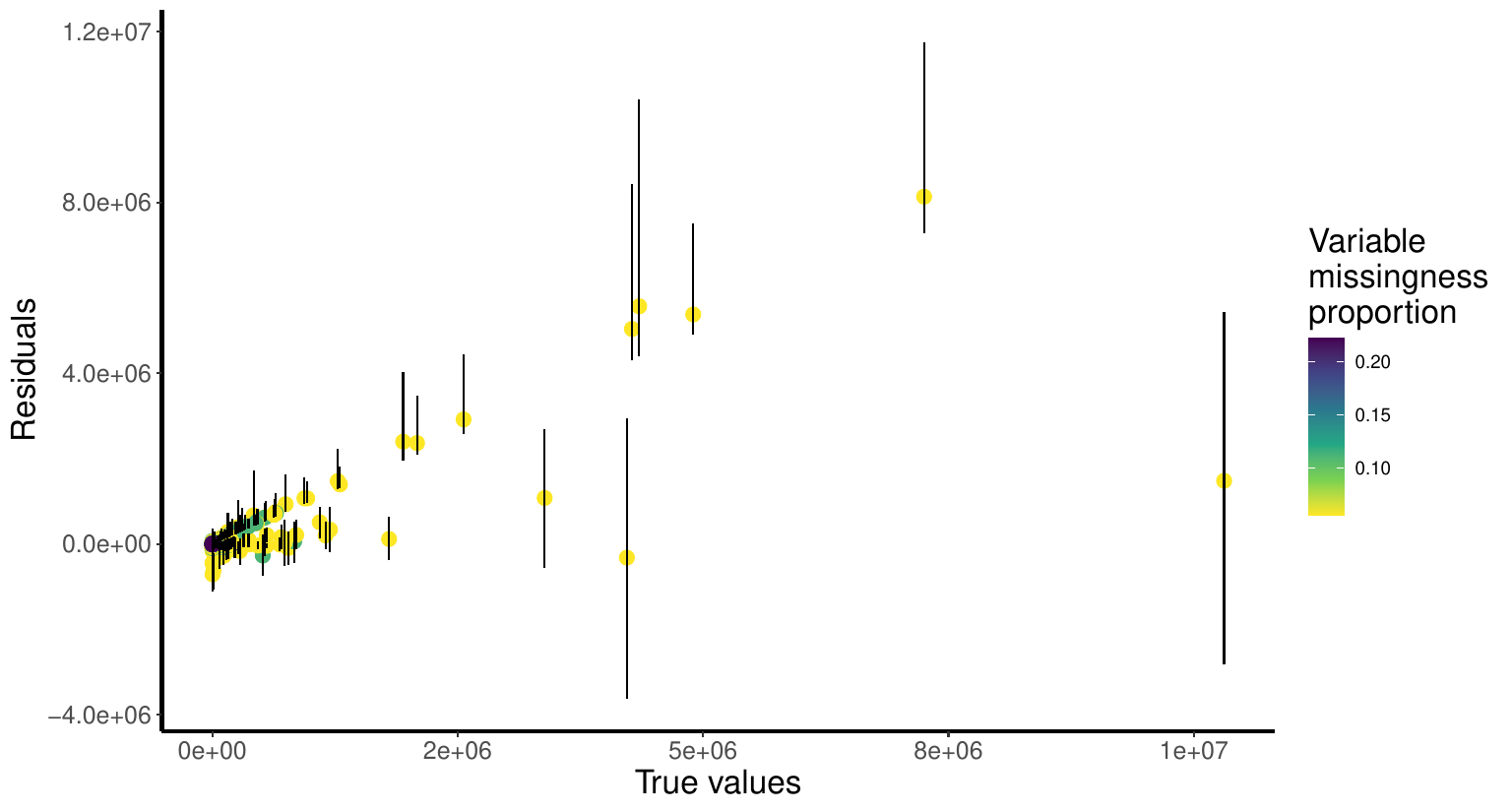}
\caption{Residuals between posterior median imputed values and true values under the IFA model for one simulated dataset.}
\label{fig:resids_ifa}
\end{figure}

\begin{figure}[htp]
  \centering
  \includegraphics[width=\textwidth, keepaspectratio]{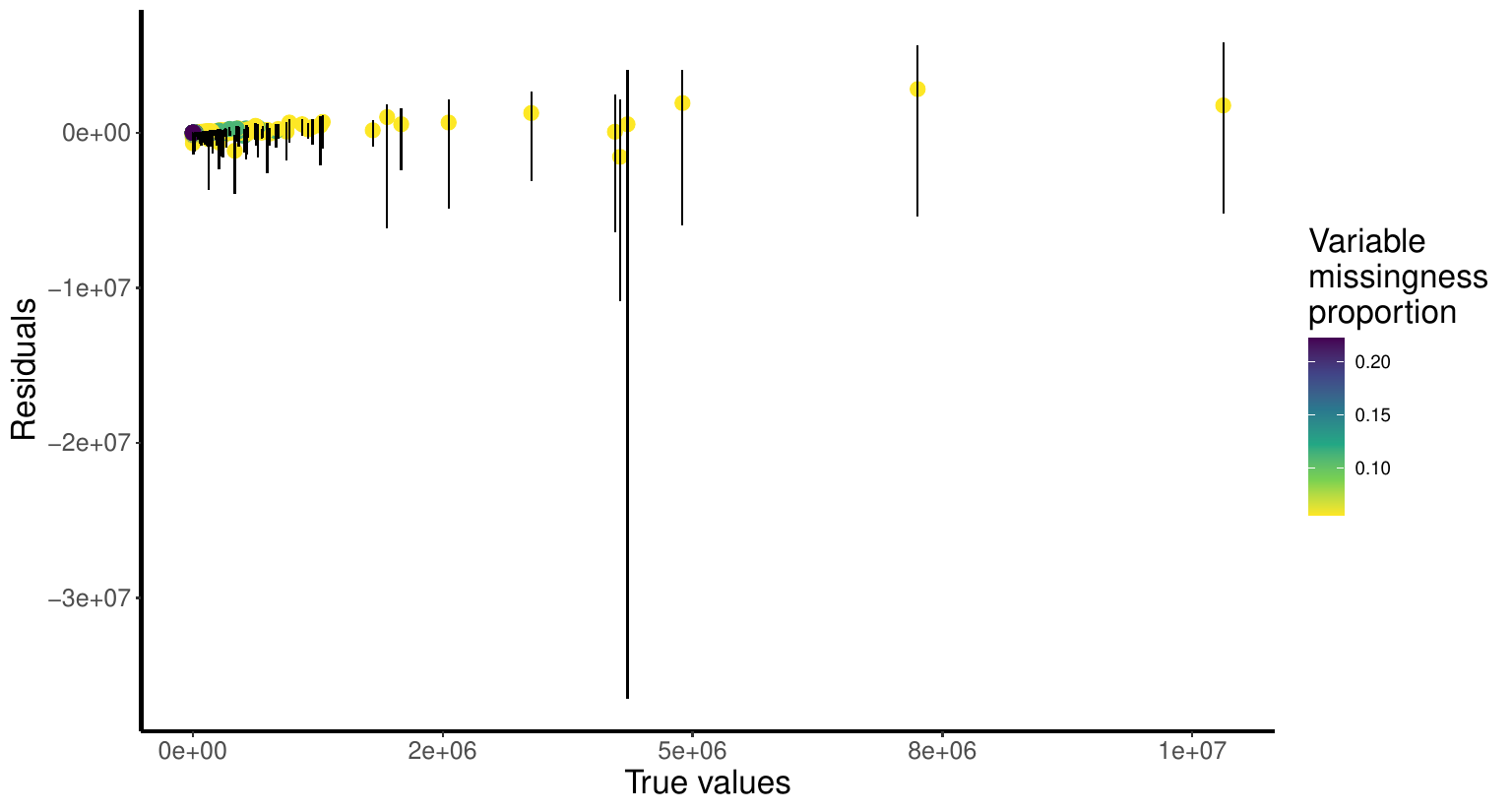}
\caption{Residuals between posterior median imputed values and true values under the IFA model applied to logged data for one simulated dataset.}
\label{fig:resids_ifa_logged}
\end{figure}

\begin{figure}[htp]
  \centering
  \includegraphics[width=\textwidth, keepaspectratio]{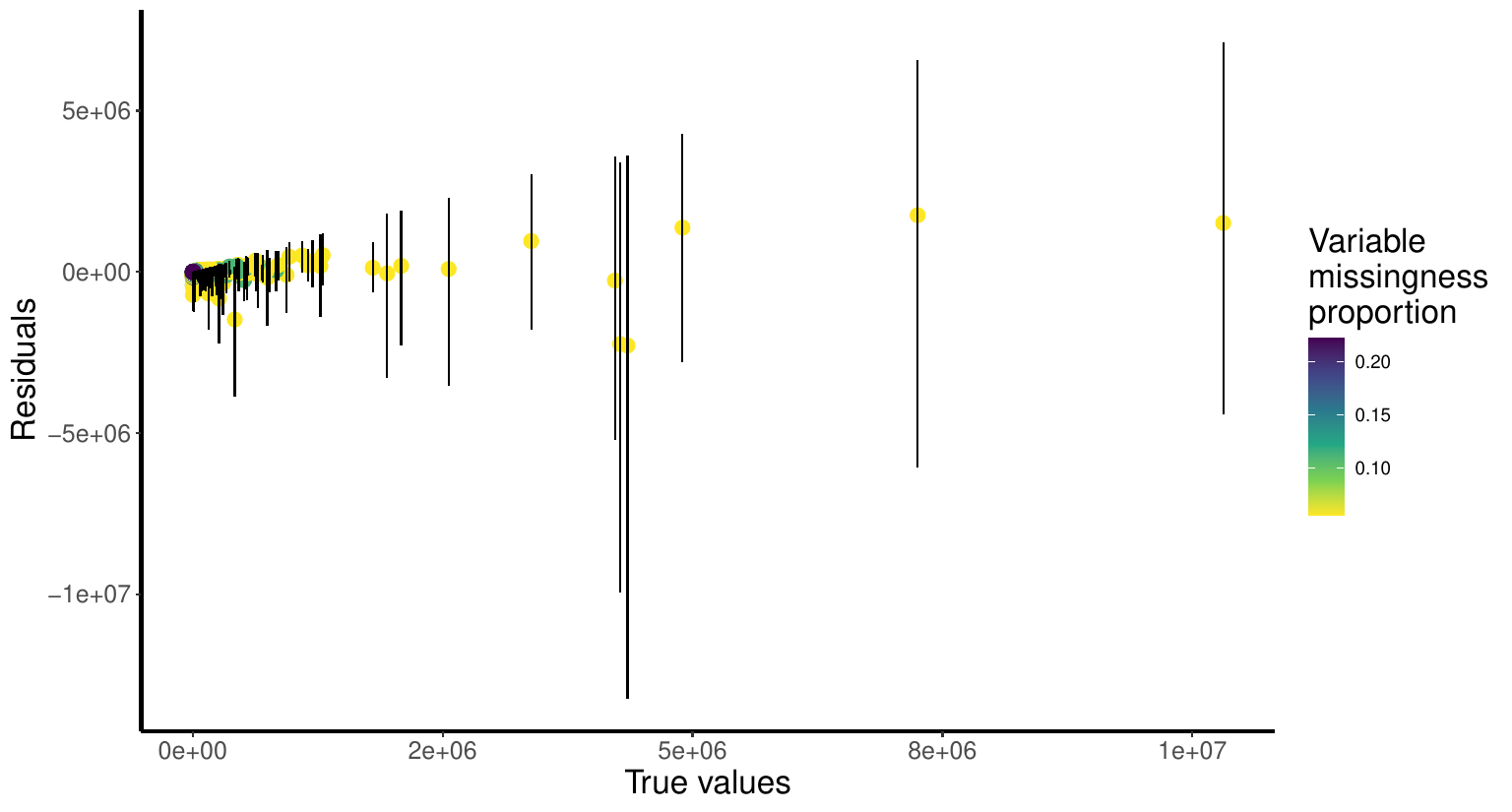}
\caption{Residuals between posterior median imputed values and true values under the TGIFA model for one simulated dataset.}
\label{fig:resids_tgifa}
\end{figure}

\begin{figure}[htp]
  \centering
  \includegraphics[width=\textwidth, keepaspectratio]{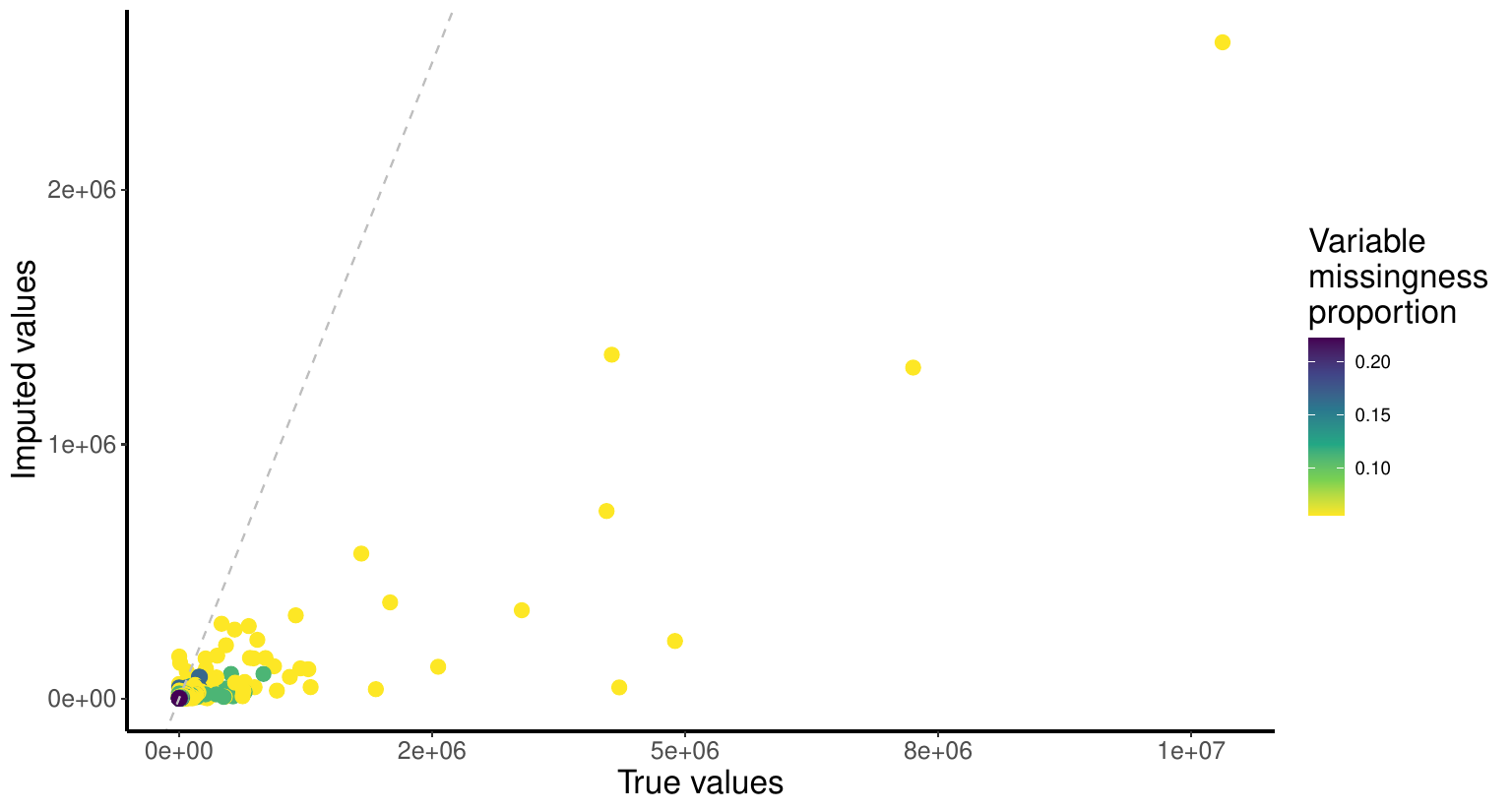}
\caption{True versus imputed values under half-minimum imputation for one simulated dataset. The dashed grey line is the line of equality.}
\label{fig:overall_imp_half_min}
\end{figure}

\begin{figure}[htp]
  \centering
  \includegraphics[width=\textwidth, keepaspectratio]{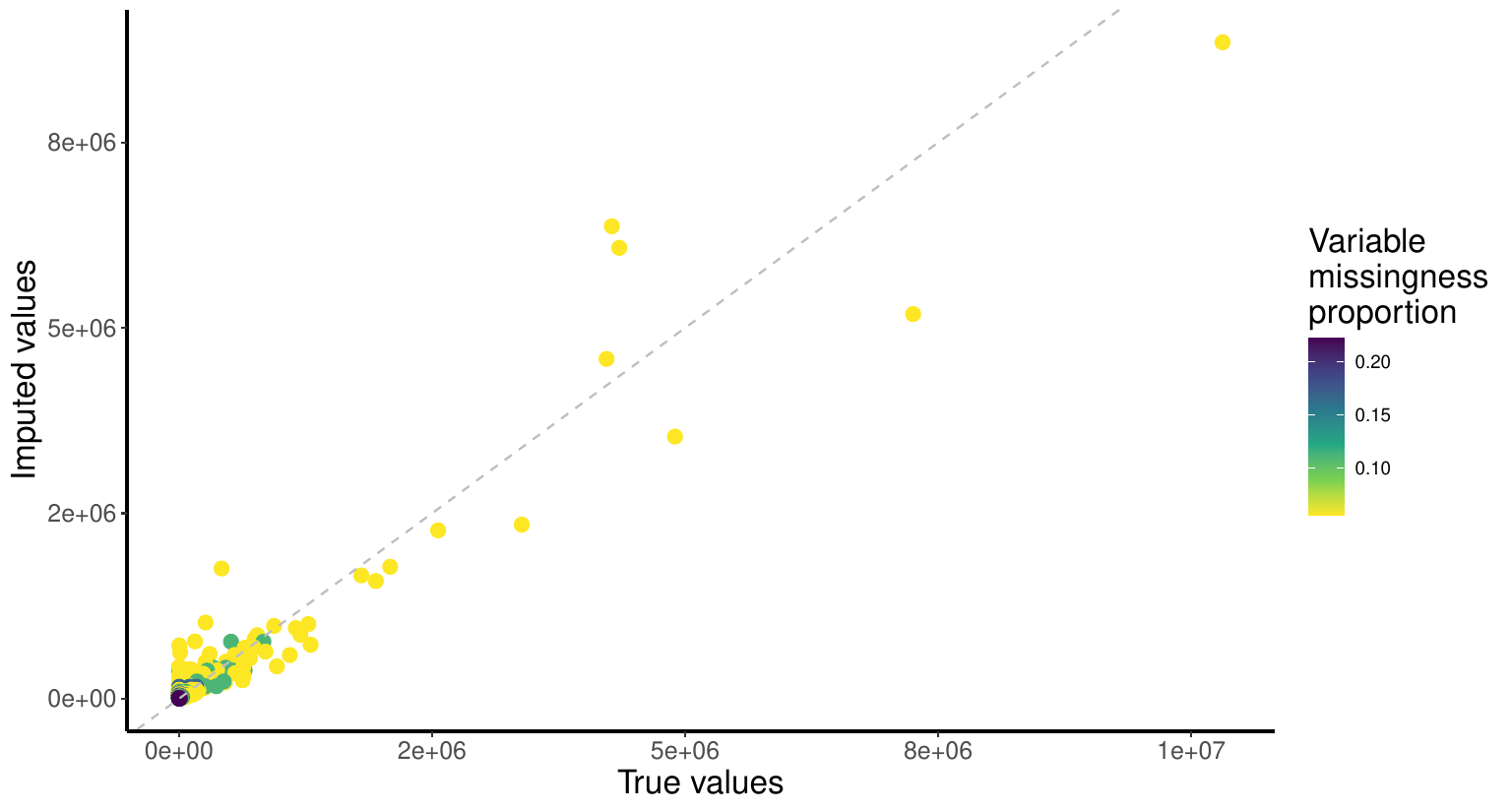}
\caption{True versus imputed values under mean imputation for one simulated dataset. The dashed grey line is the line of equality.}
\label{fig:overall_imp_mean}
\end{figure}

\begin{figure}[htp]
  \centering
  \includegraphics[width=\textwidth, keepaspectratio]{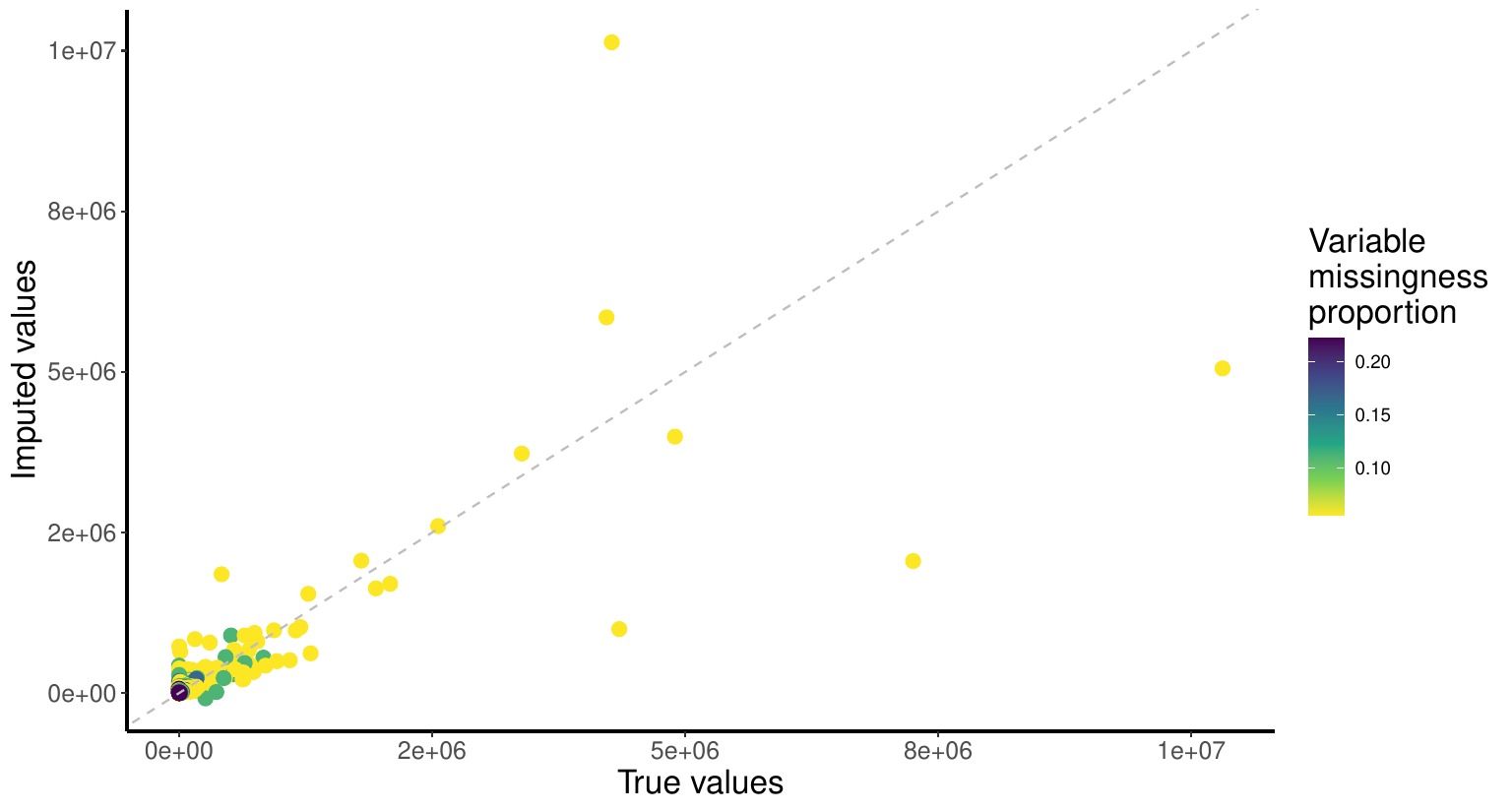}
\caption{True versus imputed values under the SVD approach for one simulated dataset. The dashed grey line is the line of equality.}
\label{fig:overall_imp_svd}
\end{figure}

\begin{figure}[htp]
  \centering
  \includegraphics[width=\textwidth, keepaspectratio]{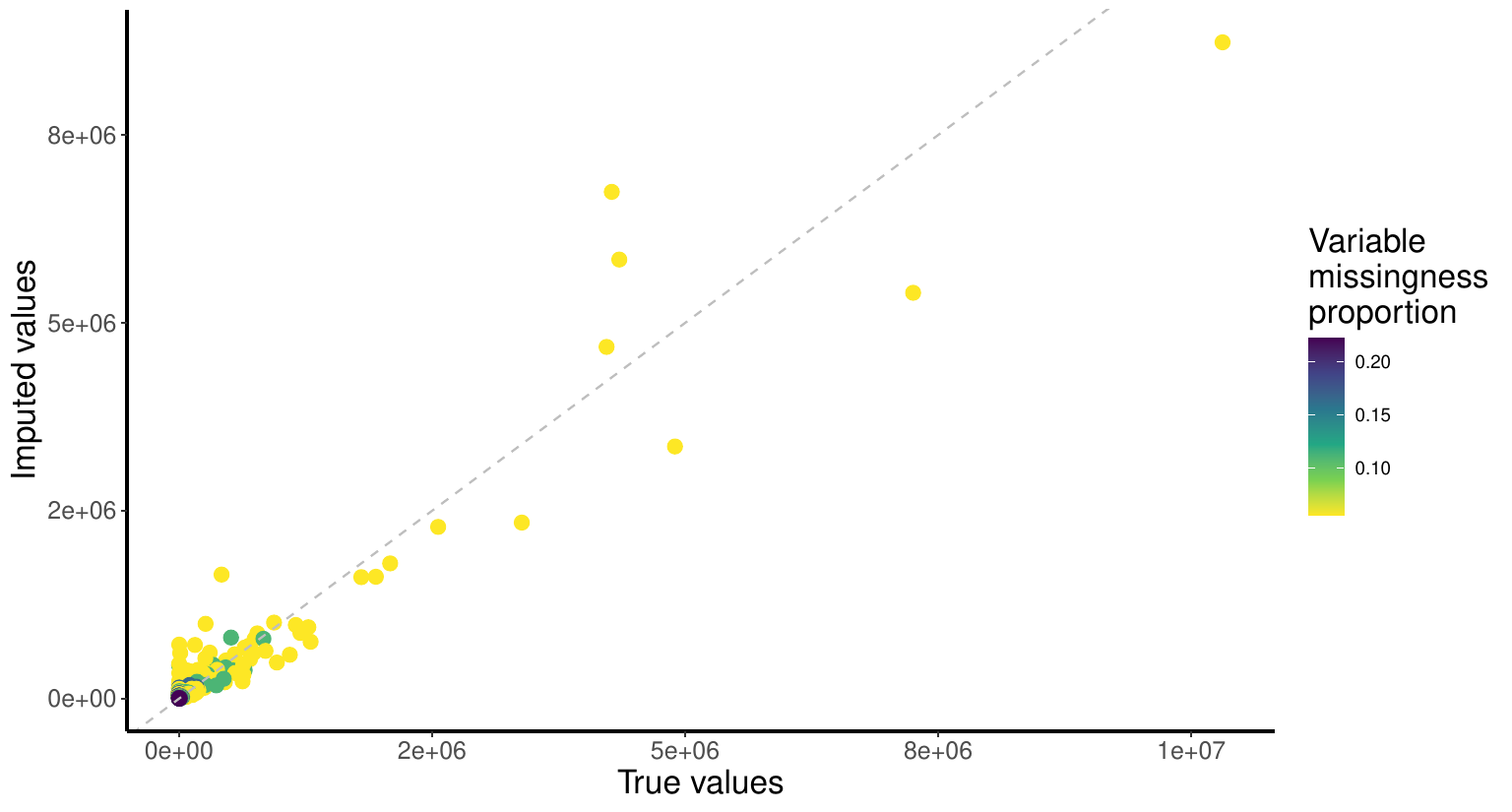}
\caption{True versus imputed values under the RF approach for one simulated dataset. The dashed grey line is the line of equality.}
\label{fig:overall_imp_rf}
\end{figure}

\begin{figure}[htp]
  \centering
  \includegraphics[width=\textwidth, keepaspectratio]{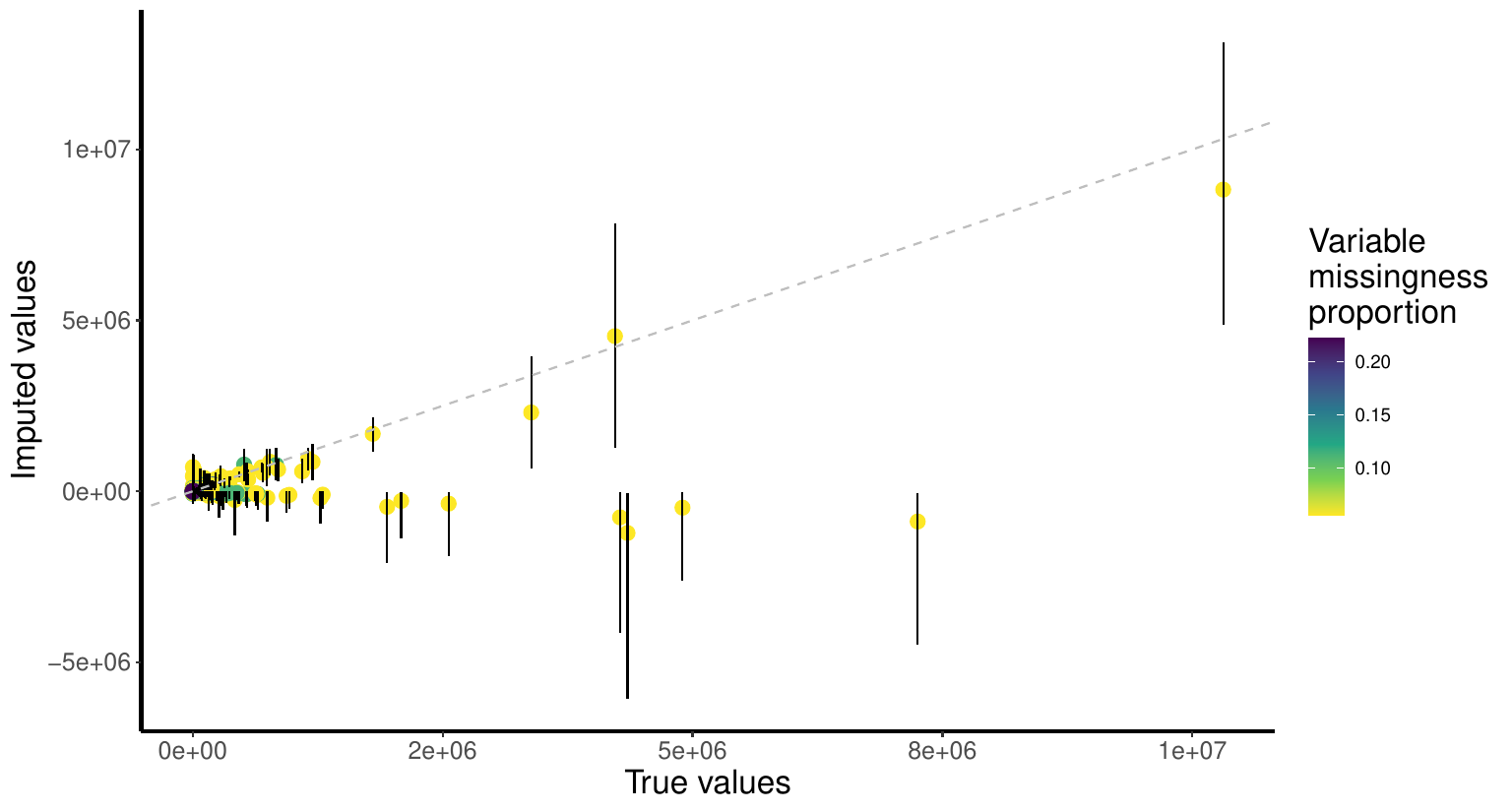}
\caption{True versus imputed values under the IFA model for one simulated dataset. The dashed grey line is the line of equality.}
\label{fig:overall_imp_ifa}
\end{figure}

\begin{figure}[htp]
  \centering
  \includegraphics[width=\textwidth, keepaspectratio]{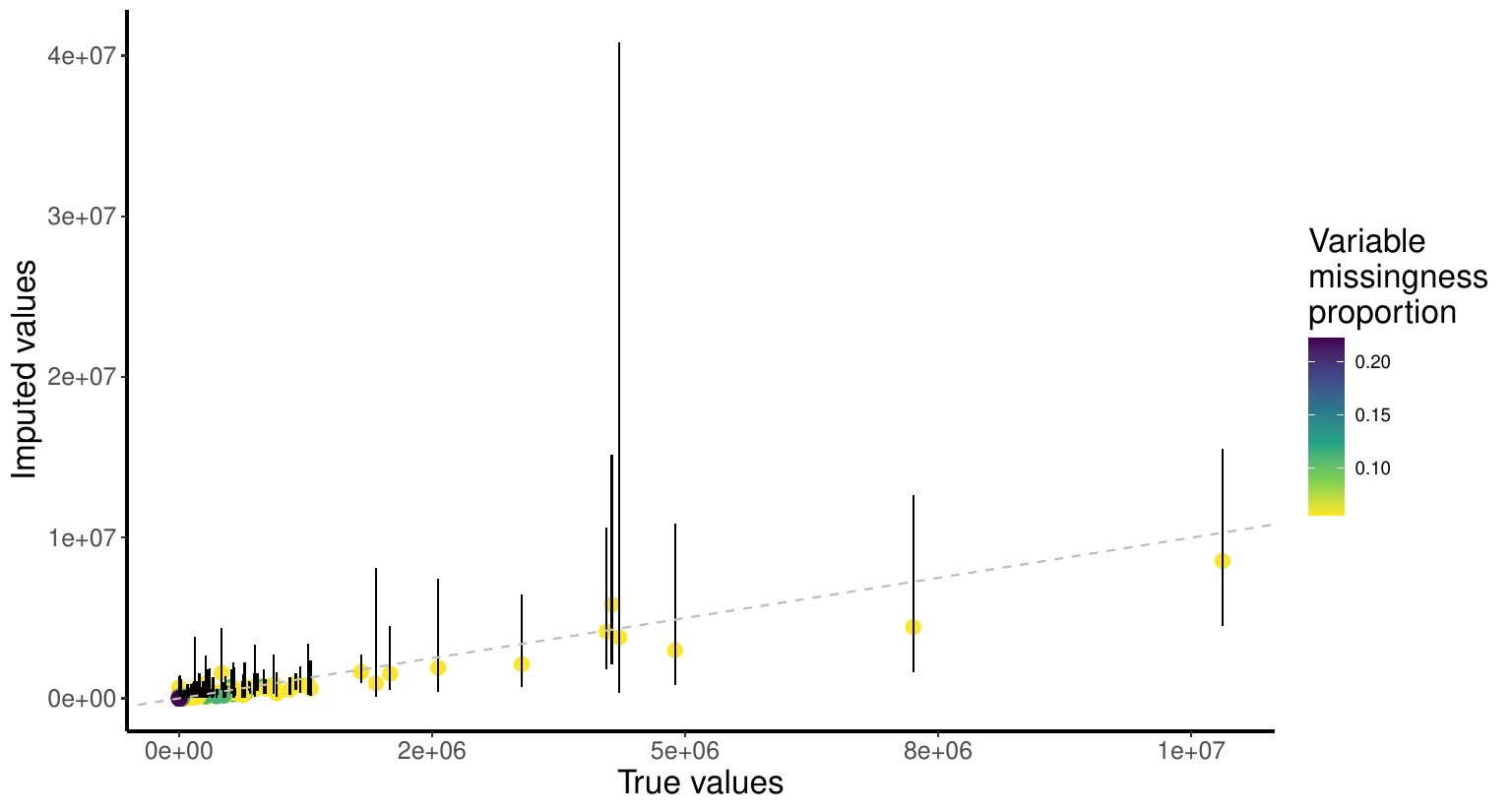}
\caption{True versus imputed values under the IFA model applied to logged data for one simulated dataset. The dashed grey line is the line of equality.}
\label{fig:overall_imp_ifa_logged}
\end{figure}

\begin{figure}[htp]
  \centering
  \includegraphics[width=\textwidth, keepaspectratio]{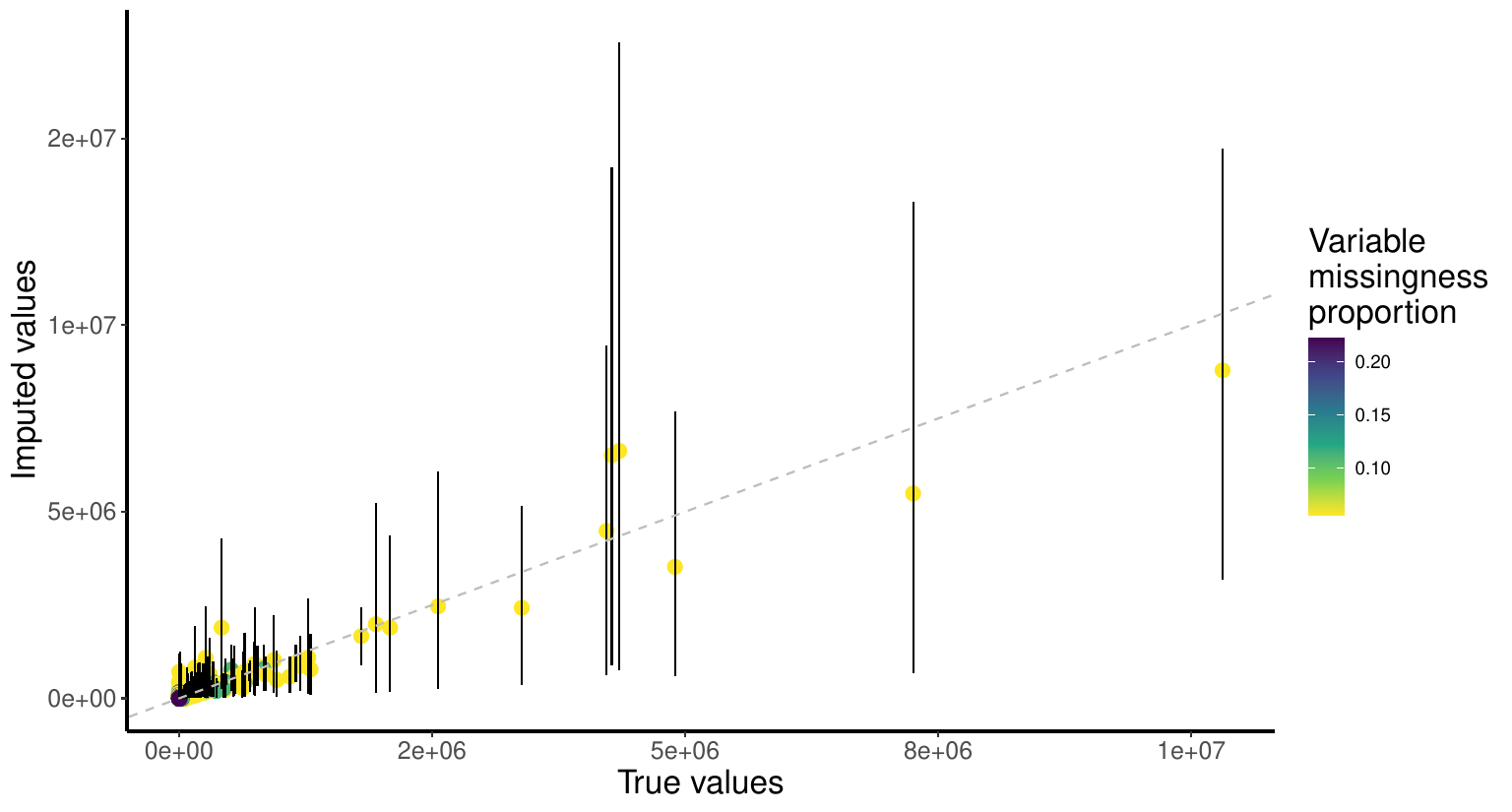}
\caption{True versus imputed values under the TGIFA model for one simulated dataset. The dashed grey line is the line of equality.}
\label{fig:overall_imp_tgifa}
\end{figure}

\begin{figure}[htp]
\centering
\begin{subfigure}{0.45\textwidth}
  \centering \includegraphics[height=0.2\textheight, keepaspectratio]{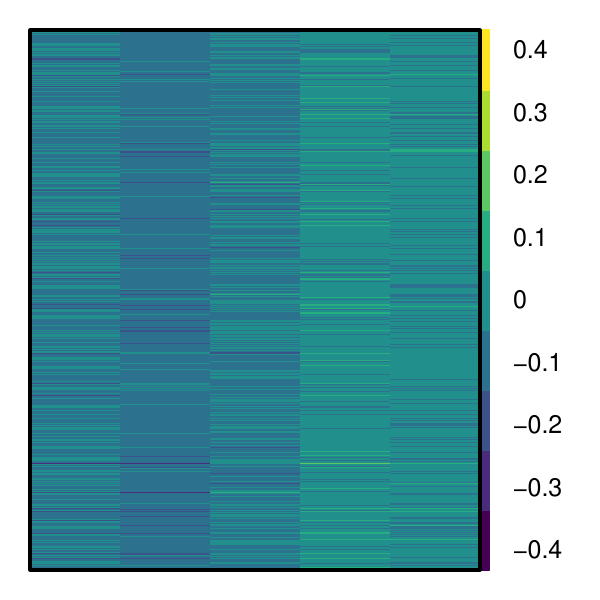}
  \caption{}
   \label{fig:loadings2}
\end{subfigure}
\begin{subfigure}{0.45\textwidth}
  \centering
  \includegraphics[height=0.2\textheight, keepaspectratio]{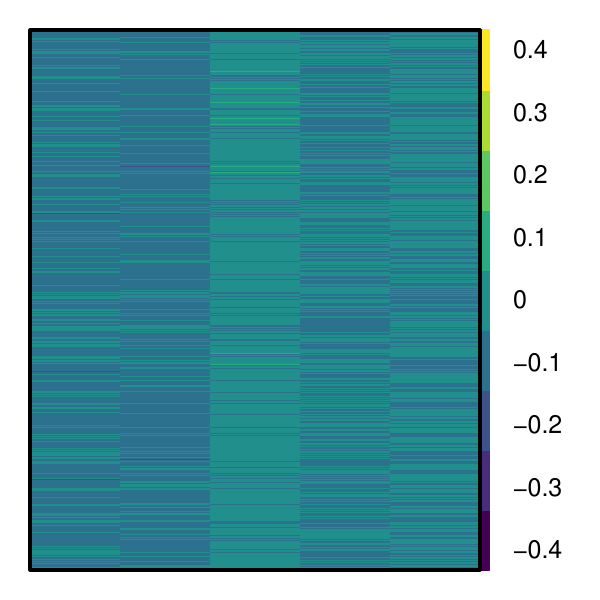}
  \caption{}
   \label{fig:loadings5}
\end{subfigure}
\begin{subfigure}{0.45\textwidth}
  \centering \includegraphics[height=0.2\textheight, keepaspectratio]{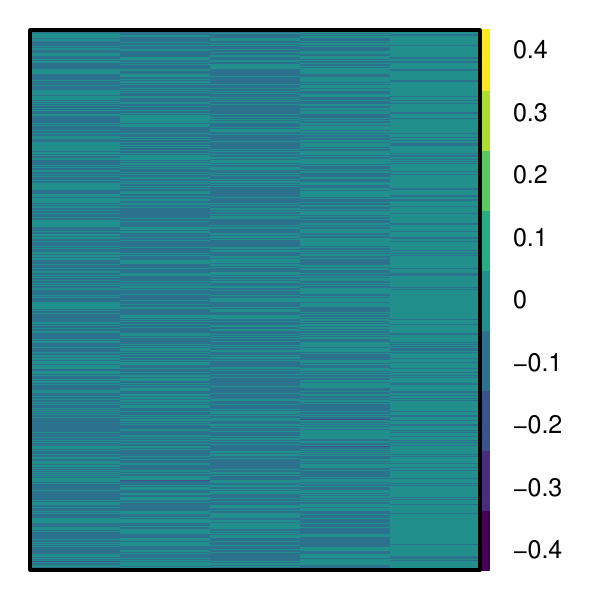}
  \caption{}
   \label{fig:loadings6}
\end{subfigure}
\begin{subfigure}{0.45\textwidth}
  \centering
  \includegraphics[height=0.2\textheight, keepaspectratio]{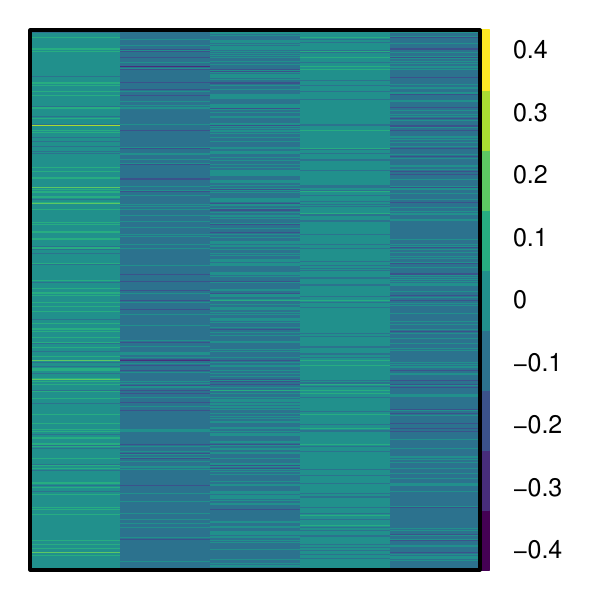}
  \caption{}
   \label{fig:loadings7}
\end{subfigure}
\begin{subfigure}{0.45\textwidth}
  \centering \includegraphics[height=0.2\textheight, keepaspectratio]{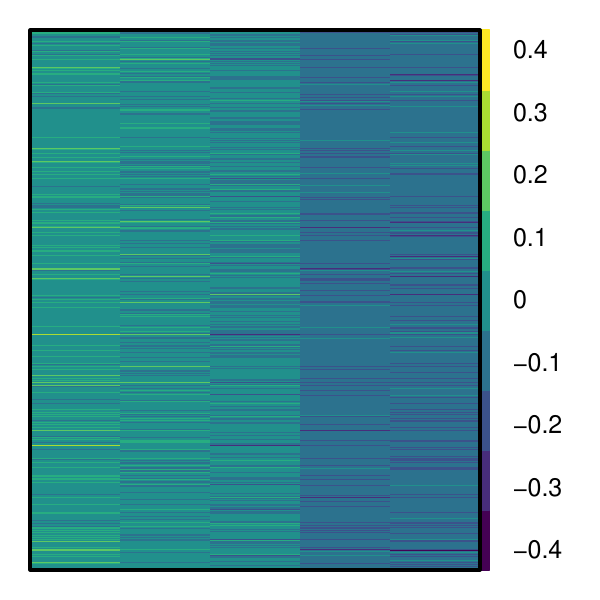}
  \caption{}
   \label{fig:loadings8}
\end{subfigure}
\begin{subfigure}{0.45\textwidth}
  \centering
  \includegraphics[height=0.2\textheight, keepaspectratio]{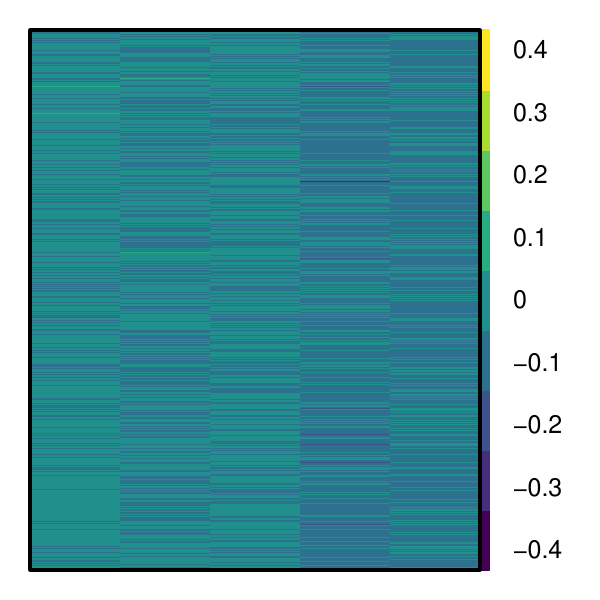}
  \caption{}
   \label{fig:loadings9}
\end{subfigure}
  \caption{The posterior mean loadings matrix under TGIFA for six simulation replicates, after Procrustes rotation with the first post-burn loadings matrix as reference in each case. Columns indicate factors. 
  }
  \label{fig:loadings}
\end{figure}

Further details of the simulation study presented in Section \ref{section:simulation} are provided here. Figure \ref{fig:resids_boxplots} illustrates the residuals between posterior median imputed values and true values under the imputation methods considered for one simulation replicate. For a single simulated dataset, Figures \ref{fig:resids_half_min}-\ref{fig:resids_tgifa} illustrate individual model residuals for half-minimum, mean, SVD, RF, IFA (applied to original and logged data), and TGIFA imputation respectively, while Figures \ref{fig:overall_imp_half_min}-\ref{fig:overall_imp_tgifa} detail the difference between the true and imputed values. Finally, Figure \ref{fig:loadings} presents posterior mean loadings matrices for a selection of simulation replicates.

\end{document}

%% file: preamble.tex
\usepackage{mathtools}
\usepackage{amsmath,amsthm,amssymb}
\usepackage{listings}
\usepackage{natbib} 
\usepackage{float}
\usepackage{mathtools}
\usepackage{ wasysym }

\usepackage[nodisplayskipstretch]{setspace}
\usepackage[dvipsnames]{xcolor}
\usepackage{graphicx}
\usepackage{subcaption}
\usepackage{mwe}
\usepackage[toc,page]{appendix}
\usepackage[nodisplayskipstretch]{setspace}
\usepackage{booktabs}
\usepackage{url}
\usepackage{lscape}
\makeatletter
\g@addto@macro{\UrlBreaks}{\UrlOrds}
\makeatother
\usepackage{enumerate}

\usepackage{pdfpages}
\usepackage[utf8]{inputenc}
\usepackage{caption}
\usepackage{multirow}

\usepackage{dsfont}
\usepackage{algpseudocode}
\usepackage{algorithm}

\usepackage[colorlinks,citecolor=blue,linkcolor=blue,urlcolor=blue]{hyperref}

\usepackage{hyperref}

\newcommand\myshade{95}
\colorlet{mylinkcolor}{violet}
\colorlet{mycitecolor}{blue}
\colorlet{myurlcolor}{MidnightBlue}

\hypersetup{
  linkcolor  = mylinkcolor!\myshade!black,
  citecolor  = mycitecolor!\myshade!black,
  urlcolor   = myurlcolor!\myshade!black,
  colorlinks = true,
}


\theoremstyle{definition}

\AtBeginEnvironment{subappendices}{%
\chapter*{Appendices}
\addcontentsline{toc}{chapter}{Appendices}
\counterwithin{figure}{section}
\counterwithin{table}{section}
}

\AtEndEnvironment{subappendices}{%
\counterwithout{figure}{section}
\counterwithout{table}{section}
}